\documentclass[sigconf,nonacm]{Chinasys/acmart}

\usepackage{array}
\usepackage{multirow}
\usepackage{float}
\usepackage{listings}
\usepackage{flafter}
\usepackage{placeins}
\usepackage{enumitem}

\graphicspath{{assets/}}
\setlist[enumerate]{topsep=2pt,itemsep=2pt,parsep=0pt,partopsep=0pt}
\definecolor{CodeBackground}{HTML}{F6F8FA}
\definecolor{CodeBorder}{HTML}{D0D7DE}
\definecolor{CodeKeyword}{HTML}{0550AE}
\definecolor{CodeType}{HTML}{116466}
\definecolor{CodeAPI}{HTML}{8250DF}
\definecolor{CodeString}{HTML}{A31515}
\definecolor{CodeComment}{HTML}{50734A}
\begin{document}

\title[uGDS: A User-Space Development Kit for High-Performance GPU Direct Storage]
  {uGDS: A User-Space Development Kit for High-Performance GPU Direct Storage}

\author[Wei et al.]{Jing Wei \quad
Guanyi Chen\textsuperscript{\textdagger} \quad
Qi Chen \quad
Chuyu Zhang\\
Zheng Zhou \quad
Chutian Wang \quad
Jian Zhang\textsuperscript{\textdaggerdbl}}
\affiliation{%
  \institution{ScaleX-IO}
  \institution{The Hong Kong University of Science and Technology (Guangzhou)}
  \country{}}

\renewcommand{\shortauthors}{Wei et al.}

\begin{abstract}
Data-intensive GPU applications, including LLM KV-cache offloading, vector search, and large-scale analytics, increasingly rely on NVMe SSDs to extend limited GPU memory capacity and support on-demand data movement. As models and datasets continue to grow beyond GPU memory capacity, efficiently transferring data between SSDs and GPUs has become critical to application performance. NVIDIA GPU Direct Storage (GDS) enables direct data transfers between NVMe SSDs and GPU memory through PCIe peer-to-peer DMA, eliminating intermediate copies through host memory. However, removing host staging alone does not eliminate software overheads or provide the flexibility required by diverse GPU workloads. Three key limitations remain: (1) the control path still traverses the kernel for every I/O request, introducing unnecessary software overhead; (2) stream-ordered I/O path incurs significant overhead and enforces a total order even when applications require ordering only at batch granularity; and (3) closed-source user-space I/O engine limits performance diagnosis and prevents applications from adapting I/O scheduling to their access patterns.

We present uGDS\footnotemark[1], an open-source user-space GPU Direct Storage
library built on four principles: user-space control, a unified abstraction
across GPU vendors, flexible workload-aware scheduling, and performance and
debuggability that both follow from keeping I/O in user space. The host CPU
constructs NVMe commands, manages hardware queues, and handles completions
entirely in user space, while payloads move directly between SSDs and GPU
memory through PCIe peer-to-peer DMA. A lightweight kernel helper
(\texttt{ugds\_drv.ko}) is used only for one-time privileged setup. Keeping
I/O initiation on the CPU matches how GPU applications already manage data
movement and consumes no GPU compute. uGDS partitions hardware queues between
synchronous and batch traffic, tracks in-flight DMA so that no buffer is
released while the SSD may still access it, and exposes cuFile-style
synchronous, batched, and stream-ordered asynchronous APIs with polling or
interrupt-driven completion on both NVIDIA CUDA and AMD HIP.

On an A100 with a Samsung 990 PRO, uGDS achieves up to 12.12$\times$ GDS's
batch-read bandwidth and 30.69$\times$ its batch-write bandwidth (8- and
4-KiB requests, batch size 128), and reduces 4-KiB synchronous write latency
by 84.1\% with two issuing threads. Its stream-ordered path sustains device
bandwidth on a single stream (5.49\,GiB/s), where GDS retains only 40.8\% of
its synchronous bandwidth. In LMCache, uGDS reduces median time-to-first-token
by up to 62.7\% and raises serving goodput by 2.63$\times$. On AMD, uGDS
achieves up to 1.19$\times$ hipFile's synchronous-write bandwidth.
\end{abstract}


% Technical report: no ACM publication metadata.
\setcopyright{none}
\settopmatter{printacmref=false, printccs=false}

\maketitle

% Keep role annotations and contact details at the bottom of the first page.
\begingroup
\renewcommand{\thefootnote}{\fnsymbol{footnote}}
\footnotetext[1]{Open-source repository: \url{https://github.com/ScaleX-IO/uGDS}}
\footnotetext[2]{Project leader.}
\footnotetext[3]{Corresponding author.}
\makeatletter
\let\@makefnmark\relax
\let\@thefnmark\relax
\let\@makefntext\noindent
\footnotetext{Email addresses (in author order):
  \href{mailto:weij63@mail2.sysu.edu.cn}{\nolinkurl{weij63@mail2.sysu.edu.cn}},
  \href{mailto:felixlinker02@gmail.com}{\nolinkurl{felixlinker02@gmail.com}},
  \href{mailto:qchen802@connect.hkust-gz.edu.cn}{\nolinkurl{qchen802@connect.hkust-gz.edu.cn}},
  \href{mailto:2200013073@stu.pku.edu.cn}{\nolinkurl{2200013073@stu.pku.edu.cn}},
  \href{mailto:zheng-zhou26@mails.tsinghua.edu.cn}{\nolinkurl{zheng-zhou26@mails.tsinghua.edu.cn}},
  \href{mailto:wangcht7@alumni.sysu.edu.cn}{\nolinkurl{wangcht7@alumni.sysu.edu.cn}}, and
  \href{mailto:jianz@hkust-gz.edu.cn}{\nolinkurl{jianz@hkust-gz.edu.cn}}.}
\makeatother
\endgroup

\section{Introduction}

Today's GPU applications face both a memory-capacity wall and a bandwidth
wall. As large language models and context windows outgrow HBM capacity,
inference engines move tensors and KV caches across GPU, host, and storage
tiers~\cite{flexgen,infinigen,llm-flash}. GPU vector search likewise relies on
efficient movement and placement of large indexes~\cite{faiss}.
Training systems such as ZeRO-Offload and ZeRO-Infinity extend GPU memory
with CPU memory and NVMe storage~\cite{zero-offload,zero-infinity};
DeepSpeed Inference combines model partitioning and offloading to serve
models beyond a single GPU's capacity~\cite{deepspeed-inference}.
These workloads make NVMe SSDs a useful intermediate storage tier. In these
workloads, the direct SSD-to-GPU transfer path lies on the critical path:
every additional microsecond of software overhead becomes GPU idle time
and can reduce compute utilization.

The conventional storage path is ill-suited to applications that move large
amounts of data between GPUs and SSDs. Requests traverse the kernel NVMe
driver, a filesystem, and often a host staging buffer, incurring system calls,
context switches, page-cache or bounce-buffer copies, and synchronization.
NVIDIA's GPU Direct Storage (GDS), exposed through cuFile~\cite{gds-overview,cufile-api},
removes host staging from the \emph{data path}: the SSD transfers data directly
to or from GPU memory through PCIe peer-to-peer DMA~\cite{pci-p2p}. Its \emph{control path},
however, still enters the kernel on each I/O: each cuFile call traverses the
filesystem stack (VFS and the underlying filesystem), the block layer, and
the kernel NVMe driver, where the \texttt{nvidia-fs} module intercepts each
request's DMA mapping and translates every 4-KiB shadow page into its cached
GPU bus address~\cite{nvidia-fs}. Large transfers amortize this fixed per-I/O cost; for small,
latency-sensitive requests prevalent in KV-cache offloading and index
lookups, it becomes the dominant cost. 

A second gap is that I/O scheduling is unaware of the ordering an application
actually needs. GPU applications express dependencies through CUDA or HIP
streams, and cuFile's stream-ordered APIs~\cite{cufile-api} inherit stream
FIFO semantics, so every request is ordered after the previous one. Many
workloads need ordering only at batch granularity: when LMCache restores a
cached prefix, its chunk reads have no order among themselves, and only the
attention kernel must wait for all of them. The stream imposes a total order
the workload does not require, and the library cannot learn otherwise. The
stream path is also costly in its own right: on a single stream, GDS retains
only 40.8\% of its synchronous 16-MiB read bandwidth, and inside LMCache, time
blocked in GDS submission calls accounts for 99.6\% of wall-clock time
(Section~\ref{sec:eval-lmcache}). Applications recover bandwidth only by
spreading I/O across many streams, in effect doing the library's scheduling
themselves. A storage library should make single-stream I/O cheap and should
schedule according to the application's real dependencies, preserving order
where it is required and exploiting the freedom where it is not.

A third gap is transparency. Mainstream solutions such as GDS do not open-source
their user-space I/O engine: although NVIDIA releases the \texttt{nvidia-fs}
kernel module~\cite{nvidia-fs}, the \texttt{libcufile} library, which
implements batch submission, worker threads, stream integration, and device
selection, is distributed only as a binary. When an application observes
stalls, such as the submission backpressure above or the first-use penalty in
Section~\ref{sec:eval-batch}, it can neither inspect where time goes nor adapt
the scheduling policy to its own access pattern.

To avoid kernel overheads, 
prior work has explored user-space kernel bypass for peer-to-peer transfers.
Research prototypes like libnvm/ssd-gpu-dma~\cite{libnvm} demonstrate the
feasibility of user-space P2P DMA between NVMe SSDs and GPUs, but remain
low-level proof-of-concept libraries: they target NVIDIA GPUs only, and their
queue primitives assume a single submitter or a lockstep group of threads
within one CUDA block, leaving concurrent submission, thread-to-queue
partitioning, and completion tracking to the application. They also do not
track in-flight commands, so a buffer can be unmapped or reclaimed while the
SSD may still DMA into it, and they offer no stream-ordered pipelining with
modern GPU runtimes. 
Industry frameworks
such as SPDK~\cite{spdk} enable high-throughput user-space NVMe control, yet
SPDK is intrinsically tailored for host-DRAM memory channels; it lacks native
awareness of GPU memory lifecycles, PCIe P2P address constraints, and
GPU stream event synchronization.

BaM~\cite{bam} takes a different route: GPU threads submit and poll NVMe
commands directly. Its performance stems from two sources: the massive thread
parallelism of the GPU, which keeps many commands in flight, and user-space
queue access, which removes the kernel from the I/O path. Only the first
requires GPU-initiated I/O, and it comes at a cost: storage polling occupies
SMs, and applications must restructure compute kernels around I/O. Meanwhile,
most GPU applications today remain CPU-centric in data management: inference
engines, KV-cache stores, and offloading runtimes decide on the host what to
move and when~\cite{vllm,lmcache,zero-infinity}. This raises the question we
address: \emph{can CPU-initiated GPU direct I/O capture the user-space benefit
and, with a carefully engineered multi-queue design on a multicore host,
deliver I/O performance competitive with GPU-initiated designs, without
consuming GPU compute or changing how applications manage data?}

uGDS answers this question with four design principles.
\textbf{User-space control:} the entire NVMe command lifecycle runs in user
space, and the kernel is involved only in one-time privileged setup.
\textbf{Unified across GPU vendors:} CUDA and HIP memory are normalized into
one controller-visible page table, so a single command engine serves both.
\textbf{Flexible, workload-aware scheduling:} because the library owns the
hardware queues, it can choose queue partitioning, completion policy,
submission coalescing, and stream integration per I/O mode, choices the kernel
block layer does not expose to applications.
\textbf{Performance and debuggability from the same structure:} keeping the
command lifecycle in one open-source user-space process both removes kernel
transitions and makes every command's state visible to ordinary debuggers,
profilers, and logs, while failures stay isolated per handle.
Together, these principles yield a clean division of responsibilities: the CPU
orchestrates, the SSD transfers, and the GPU computes. This paper makes the
following contributions:

\begin{enumerate}
  \item \textbf{User-space control path with minimal kernel scope.} NVMe
  command formatting, doorbell ringing, and completion processing execute
  directly in user space over mapped queues and registers, eliminating per-I/O
  system calls for preregistered buffers. The kernel module
  \texttt{ugds\_drv.ko} is limited to privileged setup: PCI BAR mapping, host
  DMA mapping, GPU page pinning or import, and optional MSI-X/\texttt{eventfd}
  bridging (Sections~\ref{sec:system-architecture} and~\ref{sec:device-setup}).

  \item \textbf{Concurrent, traffic-isolated queue scheduling.} uGDS partitions
  hardware queues between latency-sensitive synchronous requests and a deep
  batch queue, dispatches concurrent callers across queues without shared ring
  state, and lets applications choose busy polling or interrupt-driven
  completion (Sections~\ref{sec:queue-scheduling} and~\ref{sec:io-modes}).

  \item \textbf{DMA-safe resource lifetime.} In-flight reference counting,
  generation tokens, and fail-stop quarantine on hardware timeouts ensure that
  no buffer or queue is released while the SSD may still access it
  (Section~\ref{sec:resource-lifetime}).

  \item \textbf{Unified cross-vendor GPU memory.} CUDA and HIP/ROCm allocations
  share one controller-visible address abstraction, and registered buffers can
  be exported as DMA-bufs for RDMA and other DMA-buf-aware components
  (Section~\ref{sec:gpu-memory-abstraction}).

  \item \textbf{cuFile-style APIs with single-stream efficiency.} uGDS offers
  synchronous, batch, and stream-ordered asynchronous I/O with cuFile-like
  semantics; its stream-ordered path sustains device bandwidth on a single
  stream, so applications need not fan I/O out across many streams
  (Sections~\ref{sec:io-modes} and~\ref{sec:eval-lmcache}).

  \item \textbf{Open-source release.} The complete stack, from the kernel
  helper to the user-space I/O engine, is open source, so that users can trace
  each command, debug performance anomalies, and modify queue and completion
  policies for their workloads.
\end{enumerate}

\section{uGDS Design and Implementation}
\label{sec:design}

\subsection{System Architecture}
\label{sec:system-architecture}

\begin{figure}[!t]
  \centering
  \includegraphics[width=\columnwidth]{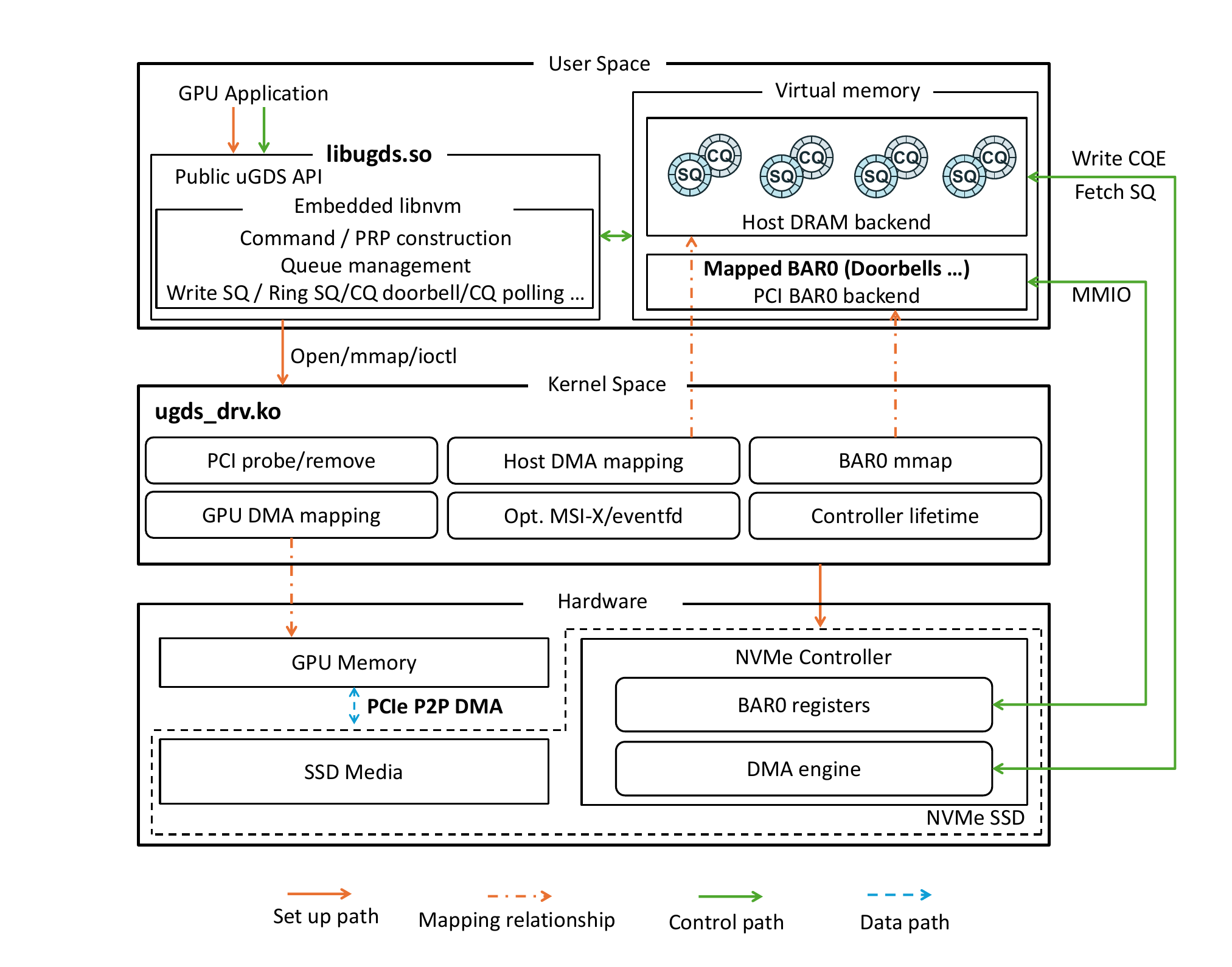}
  \caption{uGDS separates one-time privileged setup, CPU-managed NVMe
  control, and the direct SSD--GPU payload path.}
  \label{fig:ugds-architecture}
\end{figure}

uGDS decouples storage control from payload transfer to eliminate kernel
intervention while avoiding any dedicated GPU compute overhead. It enforces a
strict division of responsibilities: \emph{the CPU orchestrates, the SSD
transfers, and the GPU computes}. Figure~\ref{fig:ugds-architecture} 
illustrates the system architecture, consisting of a user-space runtime 
library (\texttt{libugds.so}), a minimal kernel helper module 
(\texttt{ugds\_drv.ko}), and the underlying hardware devices.
\vspace{5pt}

\noindent\textbf{Execution and Data Flow.}
To realize this design, uGDS organizes interactions across three distinct paths:

\begin{enumerate}
    \item \textbf{Setup Path (Privileged Kernel Delegation):} 
    To bypass the OS kernel during steady-state execution, initialization
    requires one-time privileged hardware arbitration. Applications interact with
    \texttt{ugds\_drv.ko} via standard character-device interfaces
    (\texttt{open}, \texttt{mmap}, and \texttt{ioctl}) to claim the NVMe PCI
    device~\cite{nvme-pcie}, map BAR0 controller registers and queue memory into process space,
    pin GPU virtual memory across vendors, and optionally establish MSI-X
    interrupt bridges. The driver remains entirely passive during active I/O.

    \item \textbf{Control Path (Host User-Space Orchestration):}
    During steady-state I/O, the host CPU executes the entire NVMe command
    lifecycle within user space via \texttt{libugds.so}. It formats NVMe
    submission entries (SQEs), builds Physical Region Page (PRP) lists, rings
    doorbells directly via mapped BAR0 registers, and polls completion queues
    (CQEs). By eliminating system calls, the VFS, and the block layer, the control
    overhead is drastically reduced for latency-sensitive accesses.

    \item \textbf{Data Path (Autonomous Peer-to-Peer DMA):}
    Payload data transfers occur exclusively between the NVMe SSD media and GPU
    device memory over PCIe P2P DMA. Once the SSD controller's DMA engine fetches
    the pre-pinned physical addresses published in user space, payload blocks
    bypass host DRAM, CPU caches, and bounce buffers entirely.
\end{enumerate}

\noindent\textbf{Block-Device Semantics and Abstraction Trade-offs.}
To achieve untethered user-space control over hardware queues, uGDS detaches the
target NVMe SSD from the kernel NVMe driver, operating directly over raw
block-device semantics. Consequently, applications interface with block-aligned
offsets rather than standard POSIX file descriptors. This deliberate design
choice trades hierarchical directory metadata, POSIX locks, and OS page cache
coherency for bare-metal PCIe P2P bandwidth and microsecond-level latency.

We argue that this trade-off is well-aligned with the demands of modern AI
infrastructure. Emerging high-throughput storage engines (e.g., LMCache~\cite{lmcache}
and Mooncake~\cite{mooncake}) and tiered vector indexes manage their own chunked
allocation pools, where underlying filesystem locking (e.g., in EXT4/XFS) often
introduces severe lock contention and unamortized metadata serialization. For
applications demanding file abstractions, user-space block engines like uGDS can
naturally serve as the foundation beneath lightweight user-space filesystems or
object stores (e.g., SPDK BlobFS~\cite{spdk}), maintaining near-hardware efficiency without
reintroducing kernel overheads.

The following subsections trace the lifetime of an I/O operation. Setup first
establishes device access and queue ownership
(Section~\ref{sec:device-setup}), then converts GPU memory into
controller-accessible addresses (Section~\ref{sec:gpu-memory-abstraction}).
At run time, these resources support request translation and command execution
(Section~\ref{sec:command-execution}), while queue scheduling supports
concurrent calls (Section~\ref{sec:queue-scheduling}). Resource lifetime and
failure handling span all stages to maintain DMA safety during normal
shutdown and failures (Section~\ref{sec:resource-lifetime}).

\subsection{Device Setup and Queue Ownership}
\label{sec:device-setup}

Establishing a user-space control path requires detaching hardware control
from the OS kernel and provisioning reusable queue resources. uGDS unbinds
the target NVMe PCI endpoint from the native kernel NVMe driver, enabling
exclusive user-space orchestration over controller state and queue pairs.
The application initiates device access by opening the character device
node \texttt{/dev/ugds\_drvN}; the acquired file descriptor is subsequently
encapsulated within an opaque public handle.

The kernel helper module, \texttt{ugds\_drv.ko}, manages hardware arbitration
via character-device primitives. During the PCI probe phase, it claims
BAR0, enables PCI bus mastering, configures the 64-bit DMA address mask,
and exposes the character interface. Mapping this device into process memory
exposes the controller registers with uncached memory attributes, granting
the user-space library direct MMIO access to controller registers and
doorbells. Application read/write commands are formed entirely in user
space and posted directly to the hardware engine. 
uGDS strictly follows the Linux DMA API~\cite{dma-api} to map host virtual addresses 
into device-visible bus addresses, ensuring valid physical DMA addressing.

Each file handle instantiates an isolated resource domain. All host,
CUDA, and DMA-buf mappings established via \texttt{ioctl} invocations are
tracked in a per-instance allocation ledger; overlapping ranges or
incompatible attributes are rejected at registration time. Upon file closure
or device detachment, remaining mappings are systematically reclaimed within
their respective domains. This architecture guarantees strict fault
isolation and transactional rollback during partial initialization failures.
The lifecycle management during in-flight I/O operations is elaborated in
Section~\ref{sec:resource-lifetime}.

Once hardware registers are mapped, \texttt{uGDSHandleRegister()} configures
the admin queue, identifies the controller and active namespaces, negotiates
the target number of I/O Queue Pairs (QPs), and allocates DMA-contiguous
Submission Queue (SQ) and Completion Queue (CQ) memory following the NVMe
specification~\cite{nvme-spec}. The QP allocation scales up to the maximum
capacity supported by the SSD controller. To eliminate queue contention, uGDS
statically partitions these resources: one dedicated QP is reserved exclusively
for batched I/O, while the remaining QPs concurrently serve synchronous
and stream-ordered requests. Section~\ref{sec:queue-scheduling} discusses multi-threaded queue
scheduling across these resources.

When configured for interrupt-driven completion, setup binds allocated
MSI-X vectors~\cite{msi} to kernel-bridged \texttt{eventfd} descriptors.
The kernel interrupt handler acts strictly as an asynchronous notification
shim; queue processing and CQE retirement remain under full user-space
control. In the default polling mode, worker threads check completions
directly via mapped memory. Both operational modes share an identical command
formatting, queue management, and request submission pipeline.

With controller access and queue structures in place, uGDS translates GPU
virtual allocations into controller-accessible physical addresses for
subsequent peer-to-peer data transfers.

\subsection{GPU Memory Registration and Sharing}
\label{sec:gpu-memory-abstraction}

To enable autonomous peer-to-peer DMA between the SSD and GPU, GPU virtual
allocations must be translated into bus addresses visible to the NVMe
controller. uGDS maps CUDA and HIP virtual pointers into a unified,
vendor-neutral array of controller-visible page addresses accompanied by
backend and lifetime metadata. This unified abstraction decouples the
command generation engine from vendor-specific GPU runtimes.

\begin{figure}[t]
  \centering
  \includegraphics[width=\columnwidth]{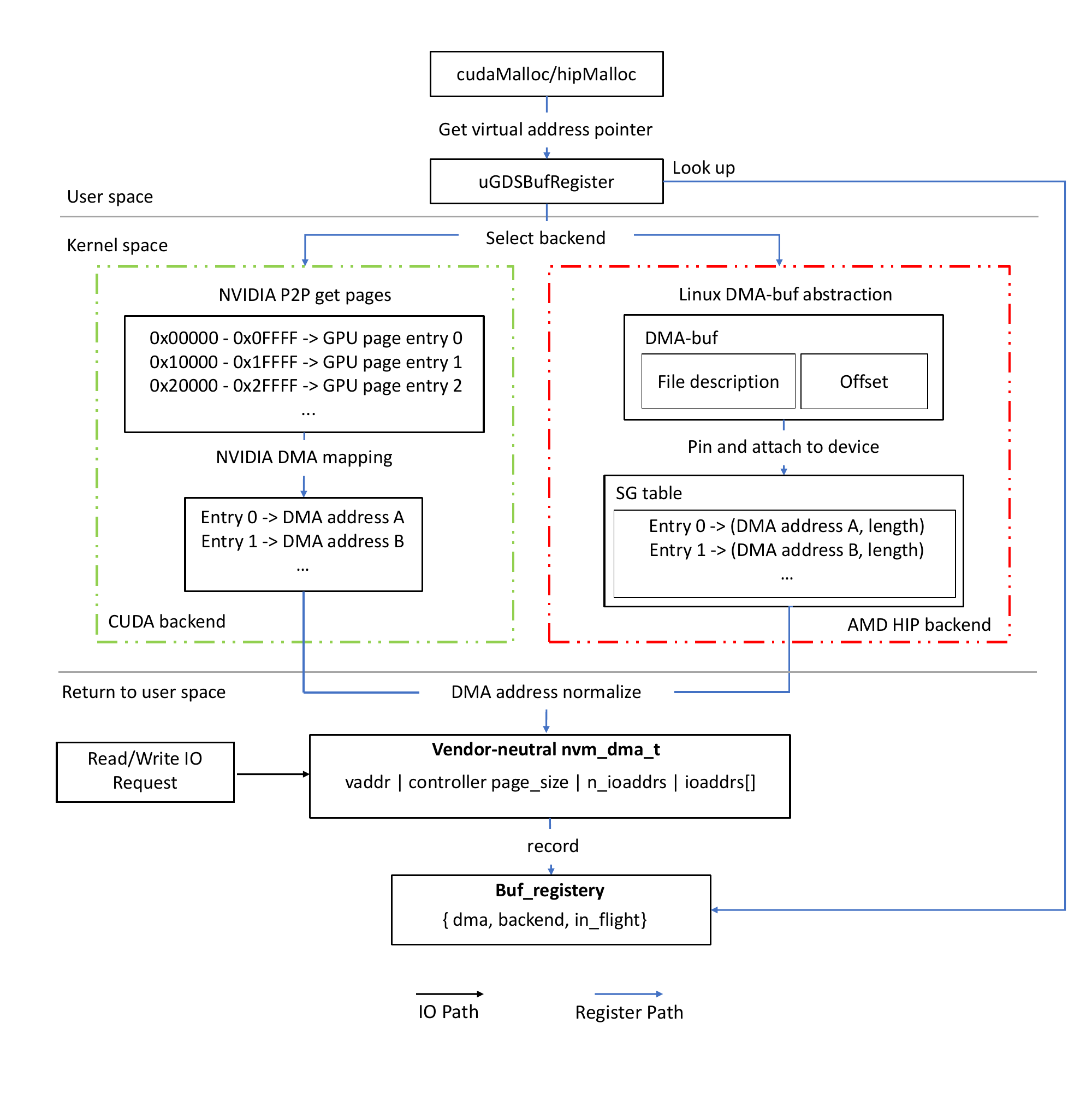}
  \caption{GPU-buffer registration converts vendor-specific GPU memory into
  one controller-visible address representation during setup.}
  \label{fig:ugds-buffer-registration}
\end{figure}

The application registers memory regions via \texttt{uGDSBufRegister()}, which
pins the underlying physical pages and records their descriptors in a
process-wide buffer registry indexed by the allocation's base virtual
address. Subsequent I/O requests reference this base pointer alongside an
explicit byte offset, preventing pointer aliasing and ambiguous lookups.
The registry continuously tracks DMA addresses, runtime backends, and
in-flight reference counts. Pinned buffers remain resident until all
referencing I/O commands retire; teardown semantics are detailed in
Section~\ref{sec:resource-lifetime}.

Figure~\ref{fig:ugds-buffer-registration} illustrates the registration
pipeline. The two runtime backends differ only in their translation
mechanisms. For NVIDIA platforms, \texttt{ugds\_drv.ko} retrieves the physical
page tables at 64-KiB granularity via the NVIDIA peer-memory
API~\cite{gpudirect-rdma} and binds them to the NVMe controller's DMA domain.
An invalidation callback coordinates with explicit deregistration paths to
guarantee safe resource reclamation. For AMD ROCm/HIP platforms, the HSA
runtime exports the target allocation as a Linux DMA-buf~\cite{dma-buf};
\texttt{ugds\_drv.ko} attaches to the DMA-buf, waits for outstanding write
fences within a bounded timeout, maps the scatter-gather (SG) list, and
serializes the entries into the unified page-address table. Consequently, 
the user-space I/O engine constructs commands and PRPs in a vendor-neutral 
manner, completely decoupled from vendor-specific GPU memory mechanisms.

Pre-registration moves pinning and translation overheads outside the critical
I/O path. While batch and stream-ordered interfaces enforce mandatory
pre-registration due to asynchronous execution lifetimes, synchronous
transfers support just-in-time (JIT) pinning for 64-KiB aligned CUDA buffers.

The registry's backend tag determines runtime-specific cleanup procedures
and asynchronous stream dispatch mechanics. Both backends share an identical
downstream storage pipeline: queue allocation, bounds verification, LBA
calculation, PRP compilation, CQE retirement, and timeout detection consume
the unified page table uniformly.

Furthermore, registered GPU buffers can be shared with external DMA
subsystems without intermediate host copies. \texttt{uGDSExportDmabuf()}
duplicates the underlying DMA-buf descriptor for external consumers, enabling
RDMA adapters to construct Memory Regions (MRs) via
\texttt{ibv\_reg\_dmabuf\_mr()}. Each registration is assigned a monotonic
generation token to prevent ABA reference hazards during buffer reuse.

By normalizing heterogeneous GPU architectures into a standard DMA
interface, uGDS establishes a consistent foundation for request translation
and command submission.

\subsection{Request Translation and Command Execution}
\label{sec:command-execution}

With mapped controller queues and pre-pinned GPU addresses, the user-space
engine translates application byte-range requests into NVMe command entries
and manages queue ring buffers. This execution pipeline forms the shared
core for synchronous, batched, and asynchronous stream operations.

Before constructing an NVMe command, uGDS validates request boundaries,
opcode semantics, buffer ranges, and target device offsets. For an NVMe
namespace formatted with logical block size $B$, the transfer parameters
must satisfy block-alignment constraints:
\begin{equation}
  \mathit{device\_offset} \bmod B = 0, \qquad
  \mathit{size} \bmod B = 0 .
  \label{eq:ugds-alignment}
\end{equation}
The translated command parameters are populated as:
\begin{equation}
  \mathit{LBA}=\frac{\mathit{device\_offset}}{B}, \qquad
  \mathit{NLB}=\frac{\mathit{size}}{B}-1,
  \label{eq:ugds-lba}
\end{equation}
where NLB denotes the zero-based logical block count encoded in command
DWORD 12. The transfer offset within the GPU buffer must adhere to the
controller's memory page size, and the span must not exceed the registered
buffer boundary.

Let $S$ represent the controller page size. For a continuous chunk of
size $C$, the required physical page descriptors are computed as:
\begin{equation}
  P=\left\lceil\frac{C}{S}\right\rceil .
  \label{eq:ugds-prp-pages}
\end{equation}
Single-page payloads are mapped via PRP1, and two-page payloads utilize PRP1
and PRP2 directly. When $P > 2$, PRP1 references the initial data page,
while PRP2 points to a host-allocated contiguous list holding the remaining
$P-1$ page addresses. Once the command and PRP structures are published,
the NVMe controller autonomously fetches payload addresses without CPU or
kernel interaction.

To bound command construction latency and memory footprint, the synchronous
path allocates one pre-mapped PRP-list page per QP. Requests exceeding
single-list addressing limits are split into multiple aligned chunks. Each
chunk is bounded by the controller's Maximum Data Transfer Size (MDTS) and the
single-page PRP capacity of $(S/8+1)S$ bytes. In the absence of an explicit
MDTS advertisement, uGDS conservatively caps transfers at 128\,KiB per chunk.
Chunk completions are accumulated in user space, reporting a single aggregated
status code upon final completion.

During command submission, the user-space library writes the 64-byte SQE
into the mapped SQ ring, increments the local tail index, and writes the
updated tail to the mapped SQ doorbell register. Sequentially consistent
memory fences ensure that PRP lists and command entries are globally visible
to PCIe before the doorbell write commits. Such strict ordering between
host memory and MMIO registers prevents race conditions in user-space
I/O engines~\cite{memory-barriers}. A single SQ slot is reserved to disambiguate
full from empty rings; larger requests are staged incrementally according
to available queue depth.

Upon receiving the doorbell, the NVMe controller fetches the commands,
executes PCIe peer-to-peer DMA against GPU memory, and writes completion
entries (CQEs) to the CQ ring. The user-space library determines completion
status by matching the hardware phase tag against the expected software cycle bit,
ensuring robust handling across ring wraparounds. Upon retiring a CQE, uGDS
validates execution status, advances the local CQ head pointer, and updates
the CQ doorbell register. Fences ensure that completion data is consumed
before freeing queue slots.

In busy-polling mode, the issuing thread polls the mapped CQ directly.
In interrupt mode, the thread suspends via \texttt{ppoll()} on the bound
\texttt{eventfd}, resuming upon notification to reap completions via the
same user-space logic. CQE entries remain the ultimate authority for
completion status regardless of notification method.

\subsection{Queue Partitioning and Concurrency}
\label{sec:queue-scheduling}

Operating directly over raw NVMe queues means software must coordinate ring
indices, doorbell registers, and in-flight descriptors without OS kernel
arbitration. Rather than letting multiple threads cooperate on a shared queue,
uGDS statically partitions the hardware QPs established in
Section~\ref{sec:device-setup} by access pattern and enforces exclusive
ownership: at any point in time, exactly one execution context may modify a
given QP.

\paragraph{Traffic-Isolated Hardware Partitioning}
Hardware QPs are divided into two functional domains: a pool of
general-purpose QPs serving synchronous and stream-ordered requests, and a
dedicated, deep QP reserved exclusively for batched I/O. This separation
prevents bursty batch submissions from inflating the queuing delay of
latency-sensitive synchronous accesses, and lets the batch QP adopt its own
queue depth and dynamic PRP pooling policy. Both domains still share the SSD
controller's internal bandwidth, but their software submission state is fully
isolated. Within its dedicated QP, the batch engine groups multiple logical
requests into coalesced doorbell submissions to amortize MMIO cost
(Section~\ref{sec:batch-io}).

\paragraph{Multi-Queue Round-Robin Dispatching}
Within the general-purpose pool, each request selects a QP through an atomic
round-robin counter and holds that QP's mutex from submission until its final
CQE is reaped. The mutex serializes all updates to the QP's SQ/CQ rings,
doorbell registers, and reusable PRP page, so ring indices and doorbell writes
need no additional atomic operations. Requests mapped to different QPs
construct commands, ring doorbells, and reap completions fully in parallel.
Because a request occupies its QP for its entire duration, a request that the
counter maps to a QP still in use waits for the earlier request to finish,
which becomes likely as the number of concurrent callers approaches the number
of synchronous QPs. This design trades intra-queue multi-producer parallelism
and per-queue depth for a simple ownership model; workloads that need deep
queues use the batch path instead.

\paragraph{Stream-Ordered Requests}
Stream-ordered asynchronous operations schedule storage transfers onto CUDA or
HIP streams via host runtime callbacks~\cite{cuda-streams, hip-streams}. Once
prior stream operations drain, the runtime executes each callback on a host
worker thread, which dispatches the transfer through the general-purpose QP
pool and waits for its completion (Section~\ref{sec:async-io}). Stream-ordered
requests therefore follow the same ownership and dispatch rules as synchronous
calls and introduce no separate queue state.

Across all execution paths, registered memory mappings are shared uniformly,
while queue state is isolated per QP. Synchronous, batched, and stream-ordered
requests can therefore run concurrently without interfering on shared queue
state, and batched I/O never contends with the synchronous pool.

\subsection{Resource Lifetime and Failure Handling}
\label{sec:resource-lifetime}
\begingroup
\setlength{\emergencystretch}{3em}

High-performance direct storage demands absolute lifecycle safety: physical
memory allocations, mapped doorbells, and queue structures must remain valid as
long as hardware DMA engines can issue transactions against their bus addresses.
Premature reclamation causes silent memory corruption or host kernel panics.
To guarantee DMA integrity across complex execution paths, uGDS coordinates
resource lifecycles through hierarchical tracking, explicit teardown ordering,
and defensive failure containment.

\paragraph{Hierarchical Reference Tracking}
uGDS governs the lifetimes of handles and buffers via centralized registries
protected by reference counts. When an application initiates an I/O, the library
atomically acquires in-flight references to both the device handle and target
buffers before command formatting. To prevent orphaned hardware operations,
child objects retain strong dependencies on their parents: a batch handle holds
an active reference to its parent NVMe handle, preventing premature controller
destruction, while exported memory descriptors retain references to the
underlying pinned GPU allocations. A global registry lock serializes reference
modifications during lookup, while decoupled per-QP locks guard queue mutation,
preserving concurrent execution across hardware queues.

\paragraph{Multi-Device Sharing and Token Validation}
When GPU memory is shared across NVMe storage, GPU compute, and external DMA
engines (Section~\ref{sec:gpu-memory-abstraction}), uGDS relies on application-level
completion barriers for producer--consumer synchronization. For example, an
NVMe read and an RDMA receive both modify the shared region and must execute
sequentially, whereas concurrent reads and RDMA transmits may overlap.
Applications enforce these boundaries using CUDA/HIP stream events or RDMA
completion tokens without embedding complex coordination logic inside the storage
library. Furthermore, to guard against address-aliasing hazards where a newly
allocated buffer reuses the virtual address of a recently freed region, uGDS
attaches a monotonically increasing generation token to each registration.
Deregistration requests must present a matching token, preventing stale or
racing requests from unpinning active mappings.

\paragraph{Orderly Teardown Semantics}
Under normal termination, applications must release resources in strict reverse
order of initialization: draining all in-flight I/O commands, revoking any
externally exported buffer references (e.g., RDMA memory regions), deregistering
GPU buffers, destroying NVMe queue handles, and finally closing the underlying
character device file descriptor. If an application attempts to release a device
handle or buffer while in-flight transfers or external references remain active,
uGDS immediately rejects the request with \texttt{-EBUSY}, preventing dangling
DMA access. To accommodate diverse application shutdown models, the extended
interface \texttt{uGDSHandleDeregisterEx()} provides explicit drain options,
supporting non-blocking status checks, bounded grace-period timeouts, and
blocking synchronization.

\paragraph{Fail-Stop Isolation on Hardware Hangs}
Hardware completion timeouts demand fundamentally stronger defensive guarantees
than typical software retry loops. When an NVMe command times out, the controller's
internal DMA engine may remain active or stalled mid-transaction. Reclaiming queue
slots or unpinning GPU memory under this condition invites delayed DMA writes
that silently corrupt application memory. Upon detecting an unrecoverable
timeout, uGDS marks the affected handle as \emph{wedged}, blocks all subsequent
submissions, and intentionally quarantines the underlying QPs and memory
references. Resource reclamation is permitted only after an external controller
reset (e.g., a PCIe Secondary Bus Reset) halts the hardware DMA engine, at which
point the application can invoke \texttt{uGDSHandleDeregisterEx()} with explicit
force flags to safely release the quarantined structures.
\par\endgroup

\section{I/O Modes and Performance Optimizations}
\label{sec:io-modes}

uGDS provides three distinct programming interfaces to cater to diverse I/O
patterns: synchronous, batched, and stream-ordered asynchronous transfers. 
This section details the operational semantics and implementation mechanics of
each interface. For brevity, subsequent code listings focus on core invocation
flows using CUDA, assuming handle \texttt{fh} and GPU buffer \texttt{gpu\_buf}
are pre-registered; AMD HIP applications follow identical semantics through
corresponding HIP runtime primitives.

Table~\ref{tab:public-api} provides an overview of the public C API exposed
via \texttt{ugds.h}.

\begin{table}[!htbp]
  \centering
  \caption{Public uGDS API.}
  \label{tab:public-api}
  \scriptsize
  \setlength{\tabcolsep}{3pt}
  \renewcommand{\arraystretch}{0.94}
  \begin{tabular}{@{}p{0.43\columnwidth}p{0.50\columnwidth}@{}}
    \toprule
    \textbf{API} & \textbf{Description} \\
    \midrule
    \texttt{uGDS\_status\_error()} & Return a string for a uGDS error code. \\
    \texttt{uGDSDriverOpen()} & Initialize the uGDS library. \\
    \texttt{uGDSDriverClose()} & Shut down the library after resources are released. \\
    \addlinespace[1pt]
    \texttt{uGDSHandleRegister()} & Create an NVMe handle for a device descriptor. \\
    \texttt{uGDSHandleDeregister()} & Drain and destroy a handle with an unbounded wait. \\
    \texttt{uGDSHandleDeregisterEx()} & Destroy a handle with an explicit drain or force policy. \\
    \texttt{uGDSBufRegister()} & Register and pin a GPU allocation for NVMe DMA. \\
    \texttt{uGDSBufRegisterEx()} & Register with explicit backend and export settings. \\
    \texttt{uGDSBufDeregister()} & Release an idle registered GPU mapping. \\
    \addlinespace[1pt]
    \texttt{uGDSExportDmabuf()} & Return a caller-owned DMA-buf descriptor duplicate. \\
    \texttt{uGDSRDMARegister()} & Create an RDMA MR for an exportable GPU buffer. \\
    \texttt{uGDSRDMAUnregister()} & Deregister the MR and release its retained export. \\
    \addlinespace[1pt]
    \texttt{uGDSRead()} & Read NVMe blocks directly into registered GPU memory. \\
    \texttt{uGDSWrite()} & Write registered GPU memory directly to NVMe. \\
    \addlinespace[1pt]
    \texttt{uGDSBatchIOSetUp()} & Create a batch object with a requested capacity. \\
    \texttt{uGDSBatchIOSubmit()} & Submit an array of logical read/write requests. \\
    \texttt{uGDSBatchIOGetStatus()} & Wait for and return batch completion events. \\
    \texttt{uGDSBatchIODestroy()} & Drain and destroy a batch object. \\
    \addlinespace[1pt]
    \texttt{uGDSReadAsync()} & Enqueue a stream-ordered NVMe-to-GPU read. \\
    \texttt{uGDSWriteAsync()} & Enqueue a stream-ordered GPU-to-NVMe write. \\
    \texttt{uGDSStreamRegister()} & Record an opaque stream without a backend hint. \\
    \texttt{uGDSStreamRegisterEx()} & Record a stream with an explicit CUDA/HIP backend. \\
    \texttt{uGDSStreamDeregister()} & Remove a stream's backend registration. \\
    \bottomrule
  \end{tabular}
\end{table}

\subsection{Synchronous I/O}
\label{sec:sync-io}

The synchronous interface exposes blocking primitives (\texttt{uGDSRead} and
\texttt{uGDSWrite}) for transferring contiguous, block-aligned byte sequences
directly between NVMe storage and GPU memory (Listing~\ref{lst:ugds-sync}).

Each invocation validates buffer boundaries and acquires an isolated hardware
queue via round-robin dispatch (Section~\ref{sec:queue-scheduling}). For transfers
exceeding single-PRP or controller limits, uGDS manages the chunked submission
loop inline according to the translation rules in Section~\ref{sec:command-execution}.
By performing inline CQE phase-tag polling on the calling core and withholding
return until all constituent chunks commit, the interface guarantees immediate
buffer reusability upon return without requiring asynchronous synchronization
barriers from the application.

\begin{lstlisting}[float=tbp,caption={Synchronous GPU--SSD I/O with uGDS.},label={lst:ugds-sync}]
uGDSDriverOpen();
int fd = open("/dev/ugds_drv0", O_RDWR);

uGDSDescr_t desc = {};
desc.type = UGDS_HANDLE_TYPE_OPAQUE_FD;
desc.handle.fd = fd;
uGDSHandle_t fh = nullptr;
uGDSHandleRegister(&fh, &desc);

void *gpu_buf = nullptr;
size_t alloc_size = 64 * 1024;
cudaMalloc(&gpu_buf, alloc_size);
uGDSBufRegister(gpu_buf, alloc_size, 0);

cudaMemset(gpu_buf, 0xab, 4096);
cudaDeviceSynchronize();
ssize_t nw = uGDSWrite(fh, gpu_buf,
                       4096, 0, 0);

cudaMemset(gpu_buf, 0, 4096);
cudaDeviceSynchronize();
ssize_t nr = uGDSRead(fh, gpu_buf,
                      4096, 0, 0);

uGDSBufDeregister(gpu_buf);
cudaFree(gpu_buf);
uGDSHandleDeregister(fh);
close(fd);
uGDSDriverClose();
\end{lstlisting}

\subsection{Batch I/O}
\label{sec:batch-io}

The batch interface decouples request submission from completion harvesting,
enabling deep I/O pipelining while amortizing MMIO doorbell overhead across
arrays of independent transfers (Listing~\ref{lst:ugds-batch}). Applications
instantiate a reusable batch context via \texttt{uGDSBatchIOSetUp()} and submit
arrays of descriptors tagged with opaque cookies, allowing out-of-order
completion harvesting via \texttt{uGDSBatchIOGetStatus()} without tracking
overhead. The batch engine executes across an isolated hardware queue pair
(Section~\ref{sec:queue-scheduling}) backed by two dedicated mechanisms:
hugepage-resident command rings and a dynamic multi-page PRP pool.

\begin{lstlisting}[float=tbp,caption={Submitting and collecting a two-request batch.},label={lst:ugds-batch}]
uGDSBatchHandle_t batch = nullptr;
uGDSBatchIOSetUp(&batch, fh, 2);

uGDSIOParams_t req[2] = {};
for (unsigned i = 0; i < 2; ++i) {
    req[i].devPtr_base = gpu_buf;
    req[i].devPtr_offset = i * 4096;
    req[i].file_offset = i * 4096;
    req[i].size = 4096;
    req[i].opcode = UGDS_READ;
    req[i].cookie = &req[i];
}
uGDSBatchIOSubmit(batch, 2, req, 0);

uGDSIOEvents_t events[2];
unsigned n_events = 2;
struct timespec timeout = {1, 0};
uGDSBatchIOGetStatus(batch, 2, &n_events,
                     events, &timeout);
for (unsigned i = 0; i < n_events; ++i) {
    if (events[i].status !=
        UGDS_BATCH_COMPLETE) {
        /* handle the per-request error */
    }
}
uGDSBatchIODestroy(batch);
\end{lstlisting}

\paragraph{Hugepage-Backed Deep Queues}
While synchronous QPs maintain an effective queue depth of one per queue due to
per-QP mutual exclusion and blocking harvest, batch mode demands deep pipelining
to maximize controller queue occupancy. uGDS provisions up to 512 queue entries
for its dedicated batch QP, far exceeding the 64-entry capacity of a single 4-KiB
page. Because the NVMe specification mandates strict physical memory contiguity
for multi-page queue rings, allocating large rings from standard fragmented pages
risks allocation failures during cold boot. To eliminate this hazard without
relying on kernel bounce buffers, uGDS allocates batch submission and completion
rings using 2-MiB hugepages via \texttt{MAP\_HUGETLB}~\cite{hugetlb}.

\paragraph{Multi-Page PRP Pool}
Sustaining deep queues introduces a severe memory hazard for transfers
exceeding two controller pages ($>8$\,KiB on 4-KiB page systems). While payloads
up to 8\,KiB embed physical bus addresses directly into the command's inline
\texttt{PRP1} and \texttt{PRP2} fields, larger transfers require an external
host-memory page to store the trailing PRP address list. In synchronous mode,
the issuing thread waits for each command to retire before initiating the next,
allowing a single PRP-list page to be safely reused across requests. In contrast,
the batch engine issues multiple concurrent commands to the SSD controller
simultaneously. Sharing a single address-list page across overlapping in-flight
commands would cause active DMA descriptors to be overwritten by subsequent
requests.

To eliminate DMA corruption without incurring dynamic memory allocation on the
critical path, \texttt{uGDSBatchIOSetUp()} provisions an isolated pool of 128
pre-mapped controller pages (512\,KiB total). Requests exceeding 8\,KiB
dynamically borrow a PRP page from this pool via an atomic bitmap, bounding the
queue to up to 128 concurrent multi-page commands (up to 256\,MiB of aggregate
in-flight data) independently of logical batch counts. Borrowed pages are
returned to the bitmap only after their corresponding CQEs retire.

\paragraph{Coalesced Submission and Incremental Reclamation}
During submission, uGDS translates each logical request into NVMe sub-commands,
maps required PRP pages, and records dependency tags in an internal tracking
table. To minimize PCIe root-complex stalls, uGDS stages commands into the SQ
ring under QP lock protection and commits the entire batch through a single
coalesced MMIO doorbell write. If hardware queue slots or PRP pages are
exhausted mid-batch, uGDS pauses command generation, rings the doorbell to
dispatch the already-staged commands, and actively reaps completions to free up
queue entries and PRP pages before staging the remaining requests. A logical
request completes once all constituent sub-commands retire, whereupon uGDS
aggregates transferred bytes, releases borrowed PRP pages to the bitmap, and
surfaces completion events.

\subsection{Stream-ordered asynchronous I/O}
\label{sec:async-io}

The stream-ordered interface allows storage transfers to be queued directly 
into CUDA or HIP streams (Listing~\ref{lst:ugds-async}), ensuring that I/O 
operations execute strictly after preceding GPU tasks and before subsequent ones.

\begin{lstlisting}[float=tbp,caption={A stream-ordered asynchronous write.},label={lst:ugds-async}]
cudaStream_t stream;
cudaStreamCreate(&stream);
uGDSStreamRegisterEx(stream,
                     UGDS_BACKEND_CUDA);

size_t size = 4096;
off_t file_offset = 0;
off_t buffer_offset = 0;
ssize_t bytes_written = 0;

cudaMemsetAsync(gpu_buf, 0xab, size, stream);
uGDSWriteAsync(fh, gpu_buf, &size,
               &file_offset, &buffer_offset,
               &bytes_written, stream);
cudaMemsetAsync(gpu_buf, 0, size, stream);

cudaStreamSynchronize(stream);
/* bytes_written now contains the result. */
uGDSStreamDeregister(stream);
cudaStreamDestroy(stream);
\end{lstlisting}

\paragraph{Host Callback Dispatch and Stream Ordering}
Asynchronous submission entry points validate handles, buffer mappings,
stream tags, and pointer parameters before queueing, incrementing reference
counts across the handle and target buffer. Depending on the backend runtime,
uGDS enqueues an internal host callback via \texttt{cudaLaunchHostFunc()} or
\texttt{hipLaunchHostFunc()} into the target stream. The GPU runtime
triggers this callback only after all upstream stream tasks complete.

Upon invocation, the callback dereferences the size and offset pointers, enabling
late-binding parameter resolution when dimensions are computed dynamically by
preceding GPU kernels. It then dispatches the request through the user-space I/O
engine and blocks until hardware CQE completion is observed, recording
transferred bytes and decrementing active reference counts. The callback blocks 
until the DMA transfer completes, preventing the GPU runtime from executing 
downstream kernels prematurely, 
strictly preserving sequential ordering within the target
stream while enabling storage transfers to overlap seamlessly with concurrent GPU 
compute tasks running on independent streams.

\subsection{Polling and MSI-X interrupt completion}

Busy polling is the default completion policy in uGDS. Applications can enable
MSI-X interrupt mode prior to handle initialization by exporting the
appropriate environment variable, as shown in Listing~\ref{lst:ugds-interrupt}.
The downstream read and write API signatures remain completely unchanged.

\noindent\begin{minipage}{\columnwidth}
\begin{lstlisting}[caption={Selecting interrupt completion before handle setup.},label={lst:ugds-interrupt}]
setenv("UGDS_INTERRUPT_MODE", "1", 1);
uGDSDriverOpen();
/* Open the device and fill desc as above. */
uGDSHandleRegister(&fh, &desc);

ssize_t nr = uGDSRead(fh, gpu_buf,
                      4096, 0, 0);
\end{lstlisting}
\end{minipage}

If MSI-X vector allocation is unavailable on a given QP, the library
gracefully reverts that queue to polling mode. In the current release,
interrupt-driven completion applies specifically to synchronous QPs; the
batch interface continuously polls its dedicated batch CQ.

In polling mode, the waiting thread continuously inspects the phase bit of
the target CQE, issuing processor \texttt{pause} instructions between cycles
to minimize bus contention without triggering kernel context switches.
In interrupt mode, handle registration queries \texttt{ugds\_drv.ko} for
available MSI-X vectors, allocates a non-blocking \texttt{eventfd} per eligible
synchronous QP, binds the vector through an \texttt{ioctl}, and configures
the hardware CQ to assert interrupts upon completion.

When an interrupt fires, \texttt{ugds\_drv.ko} increments the registered
\texttt{eventfd}. The user-space thread performs an initial speculative CQ
check, blocks on \texttt{ppoll()} with a monotonic timeout, drains the
\texttt{eventfd} counter upon wakeup, and processes new CQEs. Speculative checks
are required for two reasons. First, the \texttt{eventfd} counter does not
correspond one-to-one with CQEs: a completion may be posted before the thread
blocks, and a single interrupt may cover CQEs already consumed by an earlier
wakeup. Second, although PCIe ordering requires a CQE write to become visible
before the MSI-X write that follows it, this guarantee does not hold on every
platform; for example, Intel Meteor Lake VMD can signal an MSI before the child
device's DMA writes reach memory~\cite{vmd-mtl016}. uGDS therefore treats an
interrupt only as a wakeup hint and always re-checks the CQ phase tag. Both completion modes execute identical CQE
validation and head-doorbell updates, isolating the waiting mechanism from the
underlying storage protocol. Section~\ref{sec:eval-policy} evaluates the
resulting tradeoffs in latency and CPU utilization.

\begin{figure}[!t]
  \centering
  \includegraphics[width=\columnwidth]{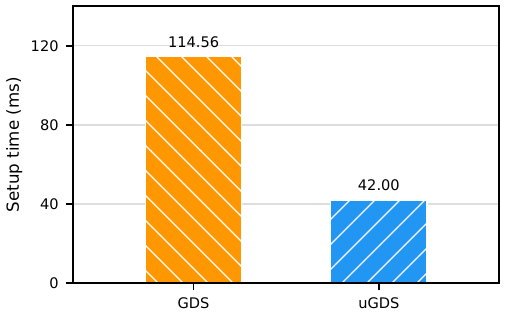}
  \caption{Application-side storage setup latency after CUDA context initialization.
  Bars represent the median of ten independent process runs (lower is better).}
  \label{fig:eval-device-setup}
\end{figure}

\section{Evaluation}
\label{sec:evaluation}

We evaluate device setup, I/O throughput and latency, and application-level
benefits in that order. Table~\ref{tab:configuration} summarizes the
microbenchmark environment; the separate LMCache configuration appears in
Table~\ref{tab:lmcache-configuration}.

\begin{table}[!t]
  \caption{Experimental configuration.}
  \label{tab:configuration}
  \centering
  \small
  \begin{tabular}{@{}>{\raggedright\arraybackslash}p{0.24\columnwidth}>{\raggedright\arraybackslash}p{0.67\columnwidth}@{}}
    \toprule
    Item & Configuration \\
    \midrule
    CPU & 2$\times$ Intel Xeon Platinum 8368Q\\
    GPU & NVIDIA A100-SXM4-40GB, PCIe Gen4 $\times$16 \\
    Storage & Samsung 990 PRO 1TB, PCIe Gen4 $\times$4; IOMMU disabled \\
    Software & Ubuntu 22.04.4, Linux 6.8.0-101, CUDA 12.4, GDS 1.9.0.20, \texttt{libcufile} 2.12 \\
    \bottomrule
  \end{tabular}
\end{table}

\subsection{Device Setup Overhead}
\label{sec:eval-setup}

Storage initialization overhead determines how long it takes for a newly 
instantiated worker process to dispatch its first storage transfer during cold 
start. We benchmark the cumulative duration of driver initialization,
device descriptor acquisition, and handle registration, starting immediately
after CUDA runtime context initialization. Both backends are evaluated across
ten independent process invocations under warm host conditions, with filesystem
metadata primed and GDS compatibility fallback disabled.

As shown in Figure~\ref{fig:eval-device-setup}, uGDS achieves a median setup
time of 42.00\,ms, compared to 114.56\,ms for GDS, a reduction of 72.56\,ms
(63.3\%). Across runs, latencies remain tightly clustered between 40.90--42.52\,ms
for uGDS and 114.23--114.69\,ms for GDS.

A stage-by-stage breakdown illuminates the structural divergence between the two
architectures. GDS consumes 111.83\,ms within \texttt{cuFileDriverOpen}, which
performs heavy initialization: it opens a session with the \texttt{nvidia-fs}
kernel driver, sets up the resources required for GDS I/O, checks platform
support, loads the \texttt{cufile.json} configuration, and discovers the PCIe
topology between GPUs and storage devices, a step NVIDIA recommends replacing
with a static topology to reduce startup time~\cite{cufile-api,gds-release-notes};
its subsequent handle registration takes only 2.72\,ms because the kernel storage
stack is already in place. uGDS distributes its setup differently:
\texttt{uGDSDriverOpen} only initializes library state and returns in 4.8\,$\mu$s,
while device-specific work is deferred to handle registration. The majority of uGDS setup latency is spent within
\texttt{uGDSHandleRegister} (41.99\,ms), which negotiates NVMe controller
registers, provisions mapped command queues, and initializes dedicated queue pairs.
During this phase, uGDS polls admin commands with a conservative 1-ms sleep
interval, intentionally sacrificing minor admin setup latency to minimize CPU spin
during cold boot. Crucially, uGDS accomplishes complete hardware resource mapping
and queue provisioning in under 42\,ms, providing a 63.3\% faster end-to-end setup
that substantially benefits elastic serving clusters frequently spawning or
recycling worker instances.

\begin{figure*}[!t]
  \centering
  \begin{minipage}{\textwidth}
    \centering
    \includegraphics[width=0.87\linewidth]{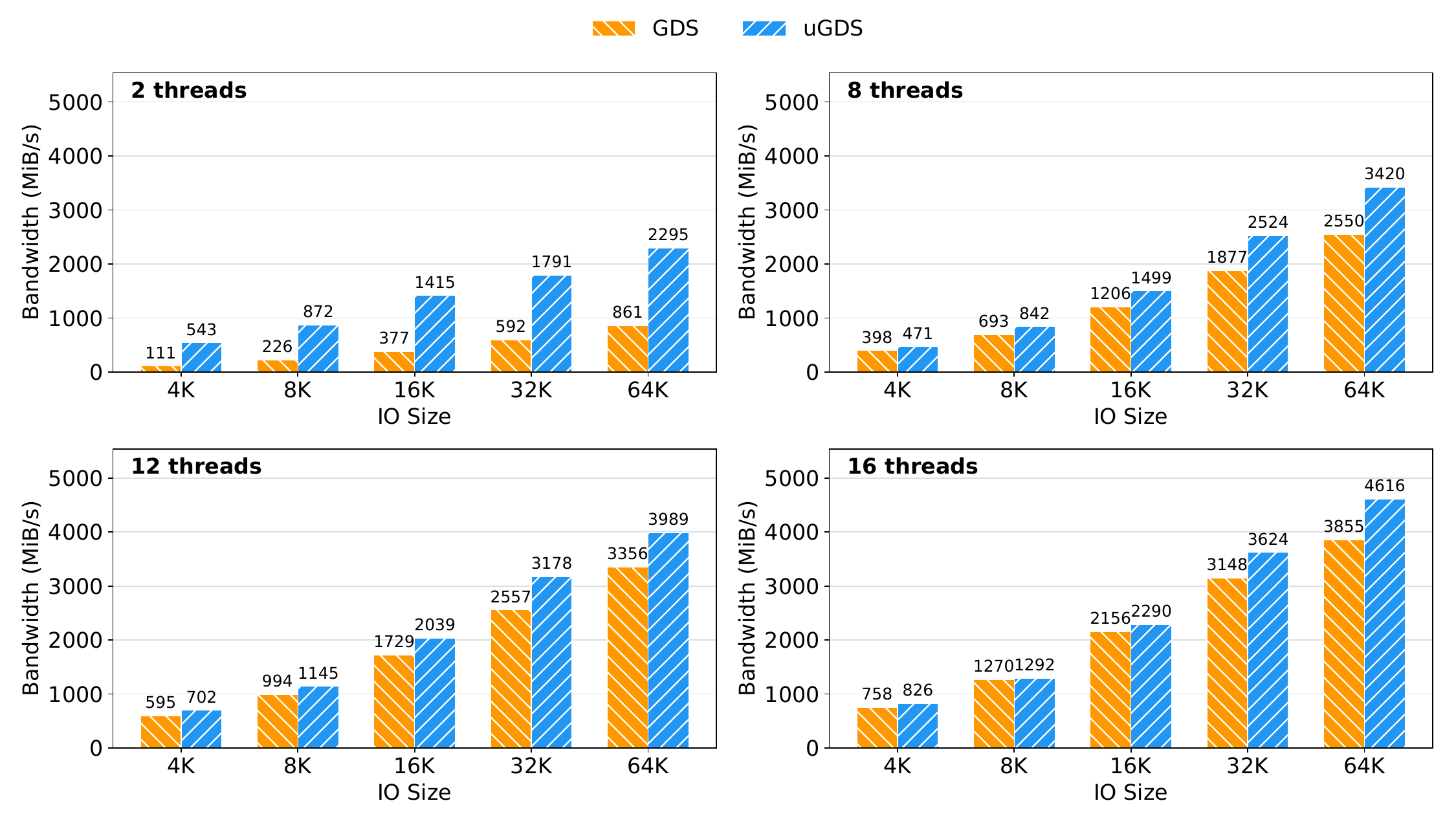}
    \captionof{figure}{Aggregate synchronous-read bandwidth with 2, 8, 12,
    and 16 issuing threads; bars are medians of three process runs.}
    \label{fig:eval-sync-read}
  \end{minipage}
  \par\vspace{5pt}
  \begin{minipage}{\textwidth}
    \centering
    \includegraphics[width=0.87\linewidth]{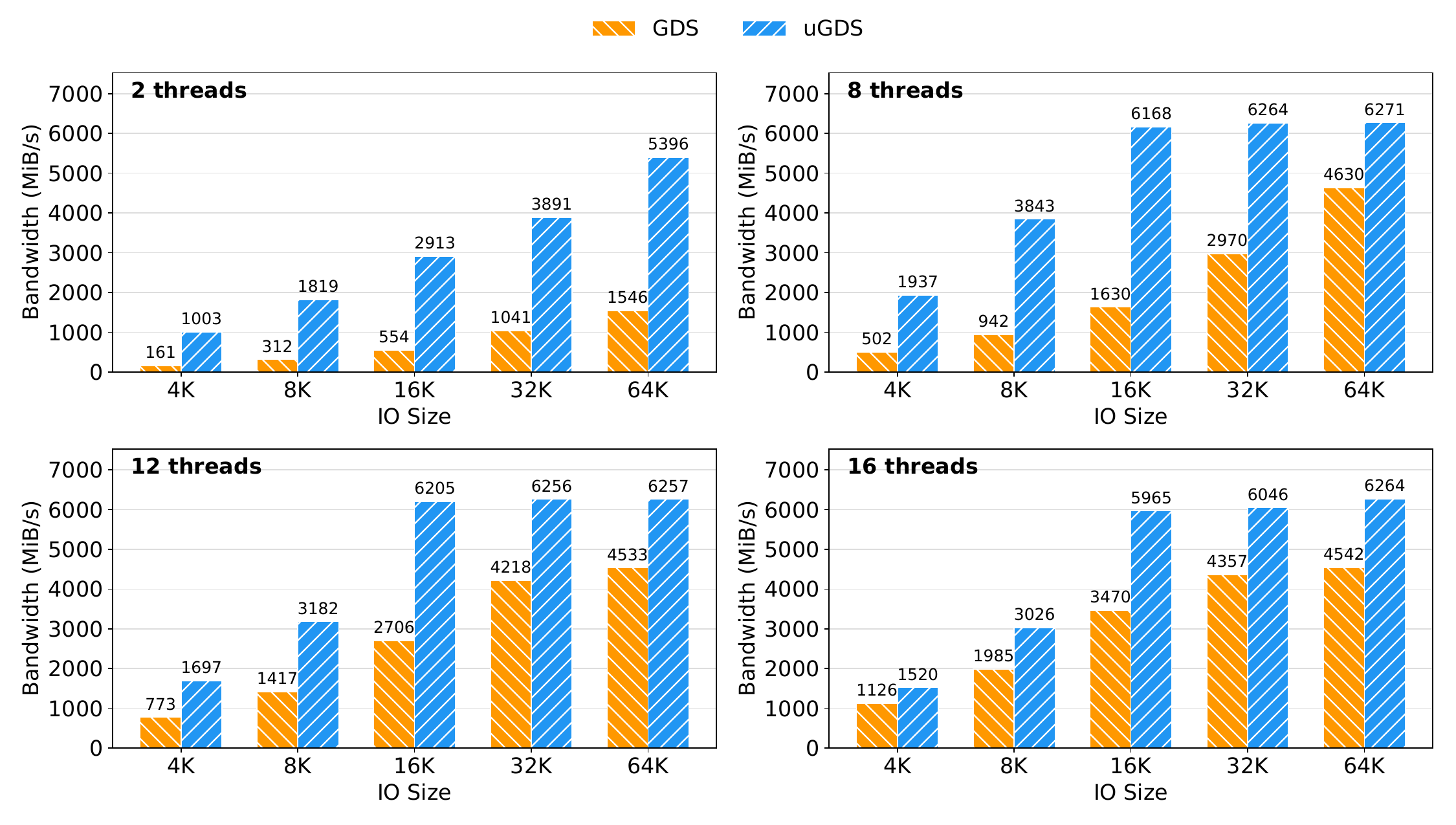}
    \captionof{figure}{Aggregate synchronous-write bandwidth with 2, 8, 12,
    and 16 issuing threads; bars are medians of three process runs.}
    \label{fig:eval-sync-write}
  \end{minipage}
\end{figure*}

\begin{figure*}[!t]
  \centering
  \begin{minipage}[t]{0.60\textwidth}
    \centering
    \includegraphics[width=\linewidth]{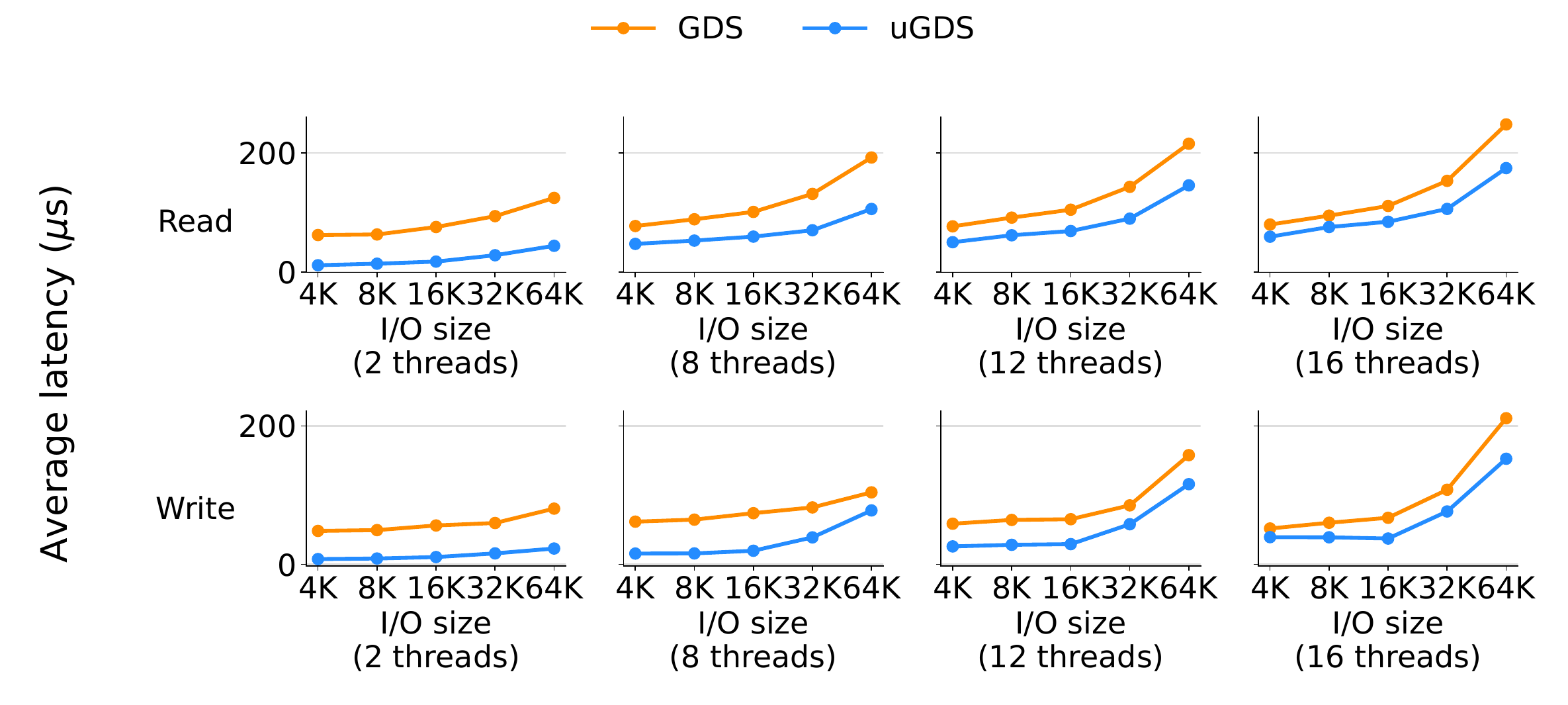}
    \captionof{figure}{Synchronous read and write latency across request sizes
    with 2, 8, 12, and 16 issuing threads. Curves report the median of three
    per-run mean request latencies.}
    \label{fig:eval-sync-latency}
  \end{minipage}\hfill
  \begin{minipage}[t]{0.37\textwidth}
    \centering
    \includegraphics[width=\linewidth]{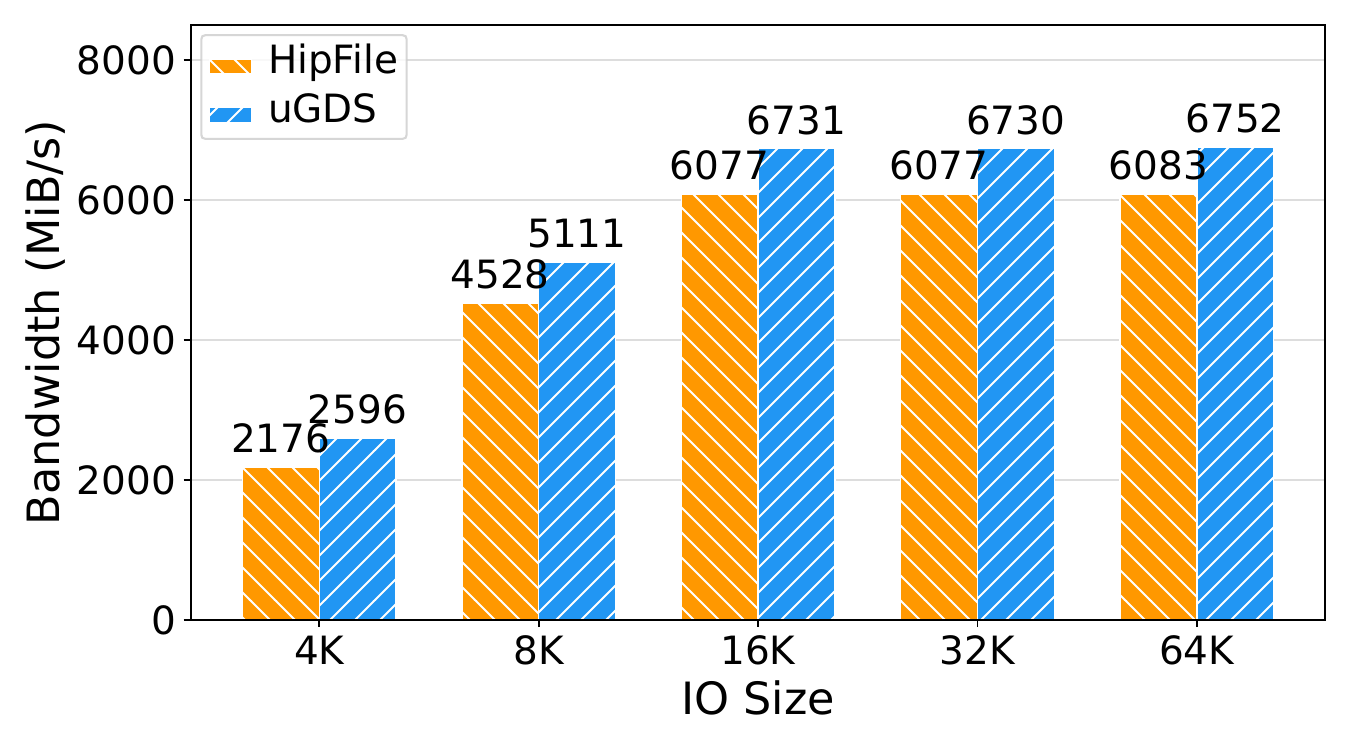}
    \captionof{figure}{Synchronous-write bandwidth on AMD with 64 issuing
    threads. Each bar is the median of three process runs.}
    \label{fig:eval-amd-sync-write}
  \end{minipage}
\end{figure*}

\subsection{Synchronous Read/Write Performance}
\label{sec:eval-sync}

We evaluate synchronous I/O on NVIDIA with 2, 8, 12, 16, 32, and 64 issuing
threads and request sizes from 4 to 64\,KiB. Each thread waits for its current
request to complete before issuing the next, maintaining an effective queue
depth of one per thread. Figures~\ref{fig:eval-sync-read}
and~\ref{fig:eval-sync-write} show the 2--16-thread range, which captures the
transition from control-path-limited execution to device-bandwidth saturation;
Figure~\ref{fig:eval-sync-latency} reports the corresponding request latencies.
Bars and latency points are medians of three formal process runs. The separate
AMD comparison uses 64 issuing threads for synchronous writes, with one
outstanding request per issuing thread.
\paragraph{Control-path-limited small I/O}
The largest synchronous gain occurs with two issuing threads and 4-KiB writes:
uGDS reaches 1003\,MiB/s versus 161\,MiB/s for GDS, a 6.23$\times$ bandwidth
increase. At the same point, the median of the three per-run mean request
latencies falls from 48.36 to 7.69\,$\mu$s, giving the maximum observed latency
speedup of 6.29$\times$ (an 84.1\% reduction) across the full 2--64-thread
sweep. Two-thread 4-KiB reads show the same pattern, with bandwidth increasing
from 111 to 543\,MiB/s (4.90$\times$). These short requests move little data and
leave device bandwidth available, so software work on each request determines
the service rate. GDS submits through the filesystem, block, and kernel NVMe
paths. For preregistered buffers, uGDS formats commands, rings mapped
doorbells, and polls mapped completion queues without a per-request system call,
which directly shortens the interval between successive transfers.

\paragraph{Concurrency and request-size scaling}
Figure~\ref{fig:eval-sync-latency} makes the concurrency transition explicit.
For 4-KiB reads, uGDS reduces latency from 62.21 to 11.56\,$\mu$s at two
threads (5.38$\times$), while the ratio narrows to 79.81 versus
59.43\,$\mu$s at 16 threads (1.34$\times$). The 4-KiB write ratio similarly
decreases from 6.29$\times$ to 1.32$\times$. Increasing the thread count exposes
more parallel requests and amortizes GDS's fixed control-path cost, while larger
requests increase the fraction of time spent transferring data. The bottleneck
therefore moves toward the SSD and PCIe path.

At eight threads and 16-KiB writes, uGDS reaches 6168\,MiB/s, compared with
1630\,MiB/s for GDS (3.78$\times$). At 16 threads and 64\,KiB, uGDS achieves
6264\,MiB/s for writes and 4616\,MiB/s for reads, versus 4542 and 3855\,MiB/s
for GDS, reducing the gains to 1.38$\times$ and 1.20$\times$. The write curves
plateau near 6.1\,GiB/s from 16\,KiB onward, whereas read bandwidth continues
to rise through 64\,KiB. The latency and bandwidth trends jointly show that
uGDS gains most from direct user-space command submission before hardware
transfer time becomes the limiting factor.

\paragraph{Cross-Platform Validation (AMD ROCm)}
To demonstrate portability across GPU architectures, we evaluate
64-thread synchronous writes on AMD hardware against hipFile~\cite{hipfile}
(Figure~\ref{fig:eval-amd-sync-write}). uGDS consistently outpaces hipFile across
all request sizes, improving 4-KiB write bandwidth from 2176 to 2596\,MiB/s
(1.19$\times$), reducing 4-KiB average API latency from 113.26 to
95.46\,$\mu$s, and sustaining 6752\,MiB/s at 64\,KiB compared to hipFile's
6083\,MiB/s (1.11$\times$). This confirms that uGDS's user-space kernel-bypass
model translates seamlessly across GPU vendors via our unified DMA-buf
abstraction.

\subsection{Batch Read/Write Performance}
\label{sec:eval-batch}

\begin{figure*}[!t]
  \centering
  \begin{minipage}{\textwidth}
    \centering
    \includegraphics[width=0.87\linewidth]{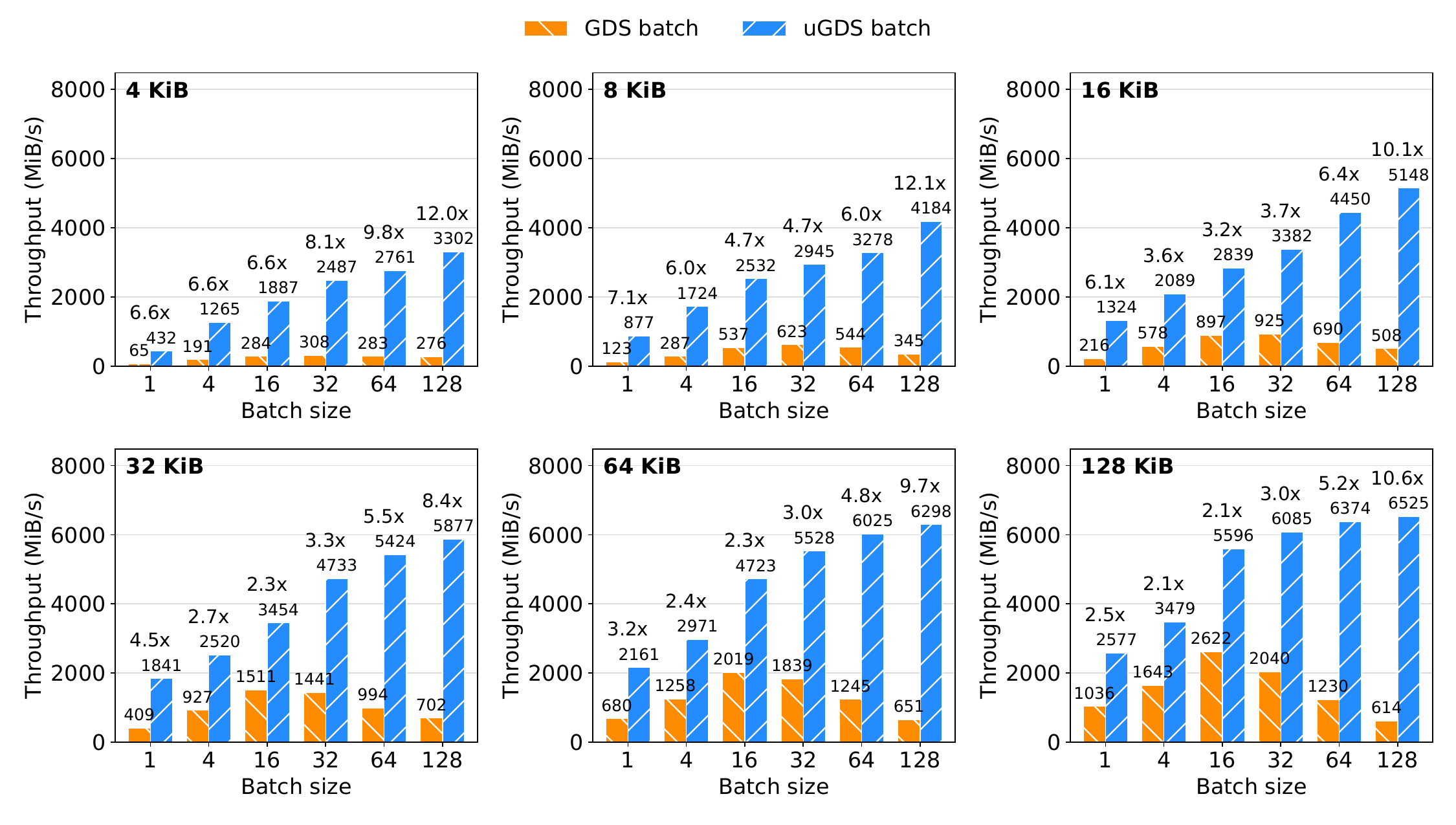}
    \captionof{figure}{Batch-read bandwidth across request and logical batch
    sizes (64\,MiB per process). Bars are medians of three runs; pair labels
    show uGDS/GDS ratios.}
    \label{fig:eval-batch}
  \end{minipage}
  \par\vspace{5pt}
  \begin{minipage}{\textwidth}
    \centering
    \includegraphics[width=0.87\linewidth]{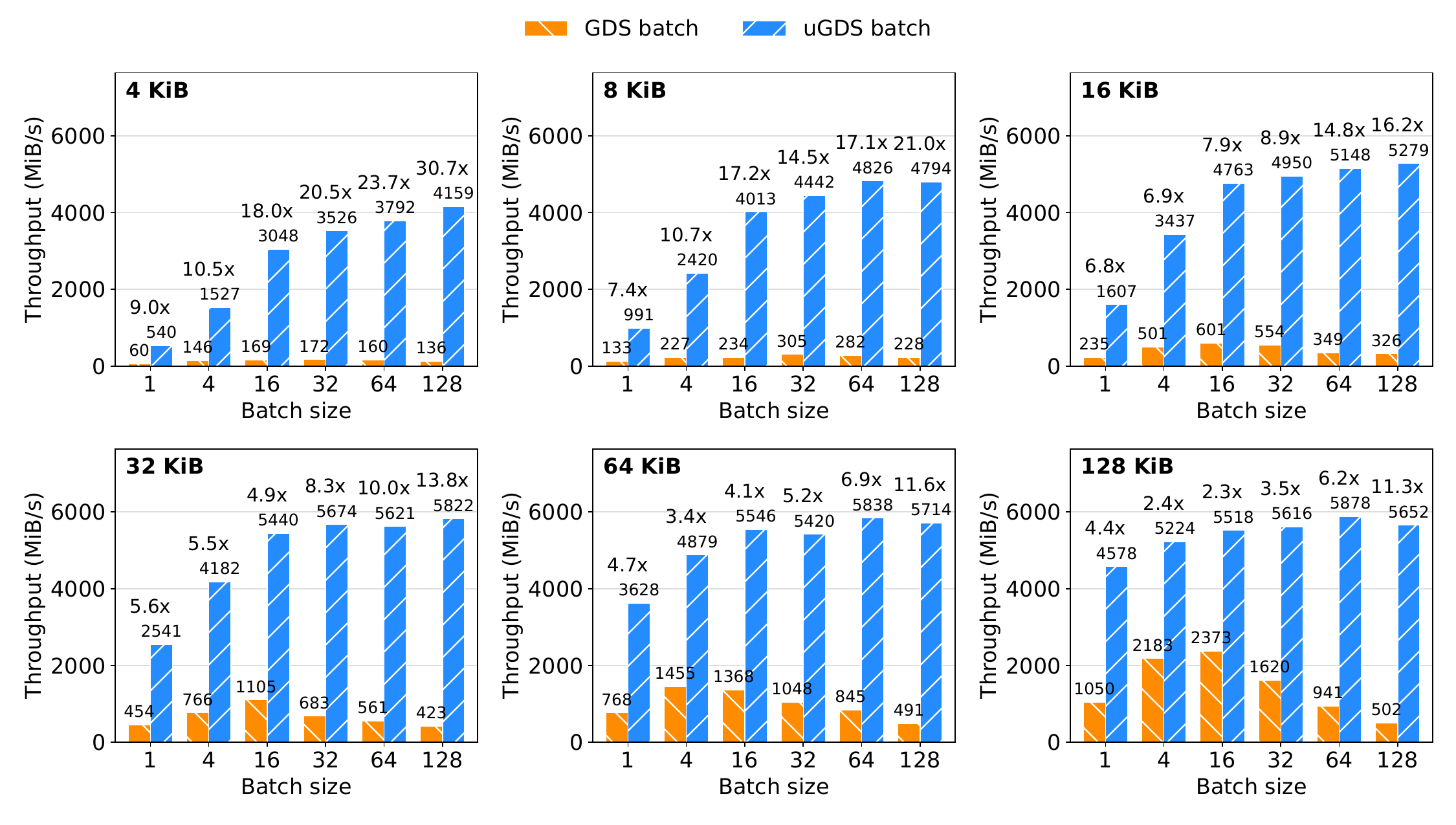}
    \captionof{figure}{Batch-write completion bandwidth across request and
    logical batch sizes (64\,MiB per process). Bars are medians of three runs;
    pair labels show uGDS/GDS ratios.}
    \label{fig:eval-batch-write}
  \end{minipage}
\end{figure*}

Batch I/O amortizes control-path overhead by queuing multiple NVMe requests
concurrently, keeping the SSD controller saturated between completion waits.
In uGDS, this path is further optimized by coalescing multiple command submissions
into a single mapped doorbell write on a dedicated deep queue. We evaluate
batch performance across request sizes from 4 to 128\,KiB and logical batch sizes
from 1 to 128. In each test, a fresh process transfers a fixed total volume of
64\,MiB by iteratively dispatching batches and awaiting completion before issuing
the next. Timings begin at the first submission call and end upon final
completion, subsequent to one-time handle registration, buffer registration,
and \texttt{BatchIOSetUp}. Each data point represents the median of three
independent fresh process invocations, deliberately capturing end-to-end
behavior including the first invocation of each process's batch path.

\paragraph{End-to-End Scaling and the GDS Anomaly.}
As illustrated in Figures~\ref{fig:eval-batch} and~\ref{fig:eval-batch-write},
uGDS comprehensively outperforms GDS across all 36 evaluated parameter
combinations, achieving 2.12$\times$--12.12$\times$ higher read bandwidth and
2.33$\times$--30.69$\times$ higher write bandwidth. For 4-KiB transfers at batch
size 128, uGDS achieves 3302\,MiB/s for reads and 4159\,MiB/s for writes, whereas
GDS yields only 276\,MiB/s and 136\,MiB/s, respectively.

More strikingly, under 128-KiB requests, GDS exhibits an anomalous performance
degradation as the batch size increases: its write throughput collapses from
2373\,MiB/s at batch size 16 to 502\,MiB/s at batch size 128. For a fixed total
volume (64\,MiB), increasing batch size from 16 to 128 reduces the total number
of issued batches from 32 down to merely 4. The catastrophic throughput drop in
GDS suggests that a severe, fixed per-batch initialization tax dominates the
total execution time when fewer batches are dispatched.

\begin{figure*}[!t]
  \centering
  \begin{minipage}[t]{0.43\textwidth}
    \centering
    \includegraphics[width=\linewidth]{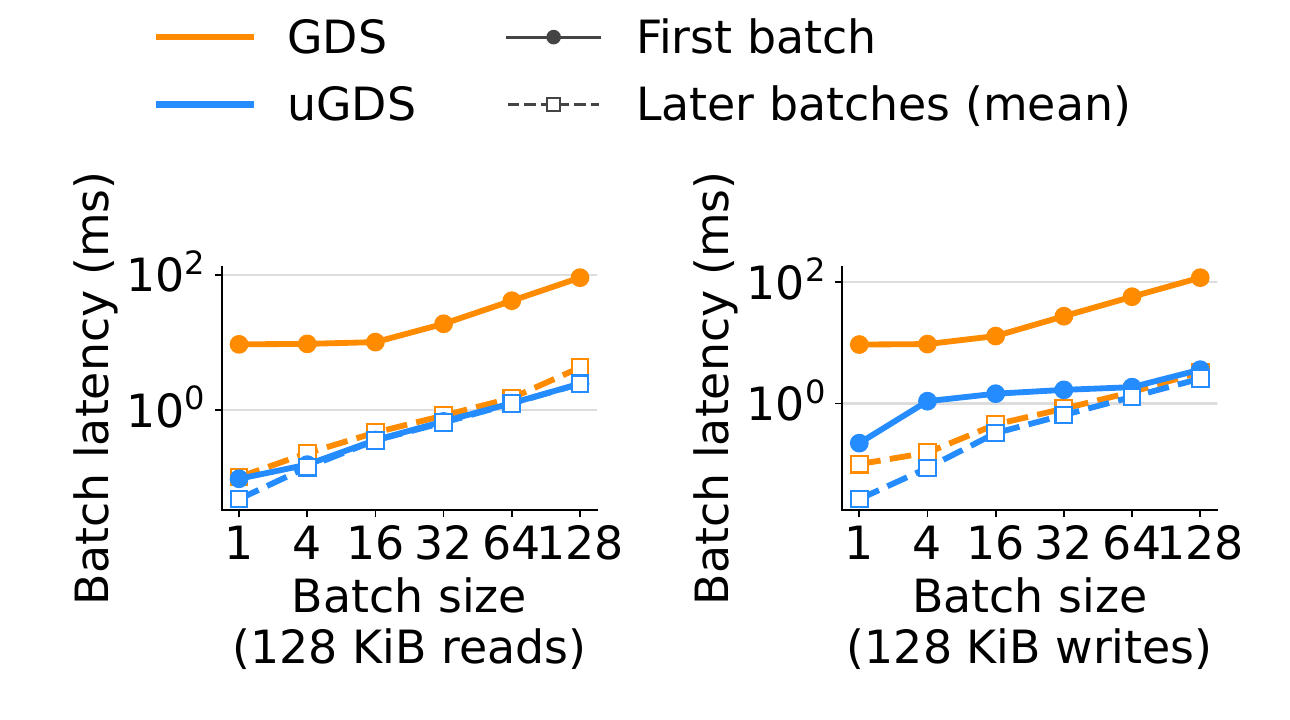}
    \captionof{figure}{Batch first-use cost for 128-KiB reads (left) and writes
    (right). Curves show first-batch latency and mean subsequent-batch latency,
    with medians over three processes and logarithmic latency axes. Setup
    precedes the timed intervals.}
    \label{fig:eval-batch-first-use}
  \end{minipage}\hfill
  \begin{minipage}[t]{0.54\textwidth}
    \centering
    \includegraphics[width=\linewidth]{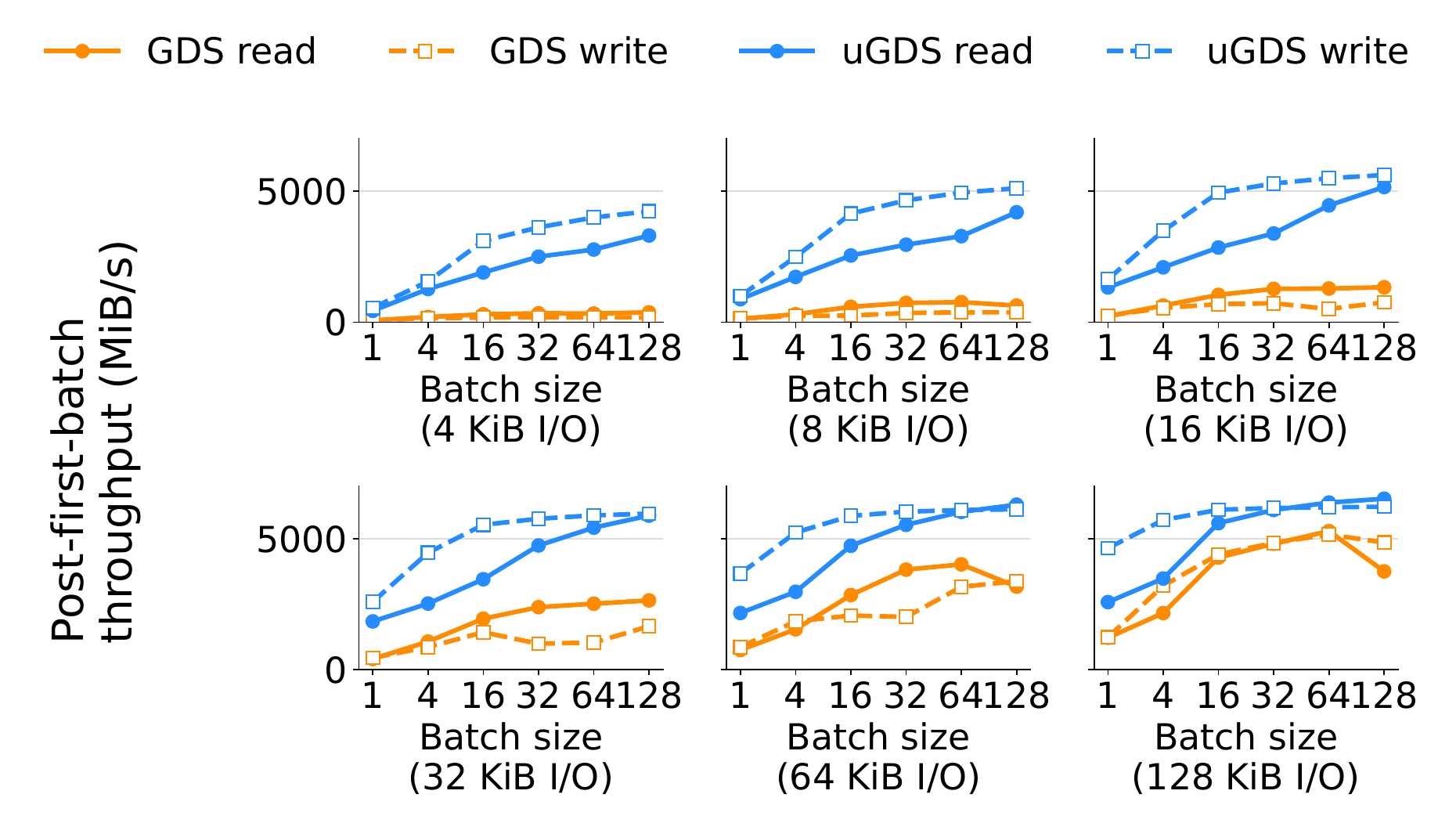}
    \captionof{figure}{Post-first-batch read and write throughput from the
    original 64-MiB transfers. Each process excludes the first batch; curves
    show medians of three processes.}
    \label{fig:eval-batch-steady}
  \end{minipage}
\end{figure*}

\paragraph{Dissecting the First-Use Penalty in GDS.}
To isolate this bottleneck, Figure~\ref{fig:eval-batch-first-use} profiles the
per-batch latency distribution for 128-KiB requests. Although explicit device
and buffer registrations are already complete before timing begins, GDS suffers
an immense latency spike on its very first batch: 90.84\,ms for reads and
117.65\,ms for writes at batch size 128. Crucially, this single first batch
accounts for 87.7\% of the total elapsed read time and 92.2\% of the total write
time across the entire 64-MiB transfer, completely crippling its aggregate
throughput. Furthermore, this initial penalty scales up proportionally with batch
size, indicating that GDS defers part of its setup to the first submission
rather than performing it during buffer registration or
\texttt{cuFileBatchIOSetUp}. A concrete instance of such deferral lies in the
\texttt{nvidia-fs} kernel module, which does not create the peer-to-peer DMA
mapping between a registered GPU buffer and an NVMe device at registration time.
Instead, it calls \texttt{nvidia\_p2p\_dma\_map\_pages()} on the first I/O
that touches that buffer--device pair and caches the result for subsequent
requests~\cite{nvidia-fs}, so the first batch absorbs a mapping cost that later
batches avoid. Because this mapping is created once per buffer--device pair, it
alone does not explain why the penalty grows with batch size; we suspect that
cuFile also lazily allocates per-request batch state, such as internal
descriptors or worker resources, sized to the first submission.

In sharp contrast, uGDS demonstrates deterministic, sub-millisecond execution
from the outset. Its first-batch read latency is identical to subsequent batches
(2.46\,ms vs. 2.45\,ms), while its write path incurs only a negligible 1.04\,ms
first-batch variance (3.61\,ms vs. 2.57\,ms). Because uGDS completes all hardware
allocations, dedicated deep queue mappings, and multi-page PRP pool bindings
upfront during explicit setup, it eliminates deferred initialization traps entirely.

\paragraph{Steady-State Throughput (Post-First-Batch).}
To verify whether uGDS's superiority is merely an artifact of GDS's first-use
penalty, Figure~\ref{fig:eval-batch-steady} evaluates steady-state throughput
by excluding the first batch's transfer size and duration from both systems.

Once this initial penalty is amortized, GDS recovers substantial bandwidth for
large transfers: at 128\,KiB with batch size 128, GDS reaches 3755\,MiB/s (reads)
and 4855\,MiB/s (writes). Nevertheless, uGDS still maintains a decisive lead,
delivering 6530\,MiB/s (1.74$\times$) and 6223\,MiB/s (1.28$\times$), approaching
the hardware bandwidth ceiling.

For fine-grained, latency-sensitive requests (e.g., 4\,KiB), uGDS's performance
advantage remains overwhelming even in pure steady-state operation: uGDS achieves
3308\,MiB/s for reads and 4232\,MiB/s for writes at batch size 128, outperforming
GDS (374 and 178\,MiB/s) by 8.8$\times$ and 23.8$\times$. This demonstrates that
GDS's recurring per-batch kernel mediation imposes severe recurring costs on small
I/O. In contrast, uGDS manages request arrays, populates PRPs from its preallocated
host-mapped pool, rings doorbells once per batch, and harvests CQ completions
entirely within user space, making it uniquely suited for high-frequency,
bursty batch workloads common in LLM serving.

\begin{figure}[!t]
  \centering
  \includegraphics[width=\columnwidth]{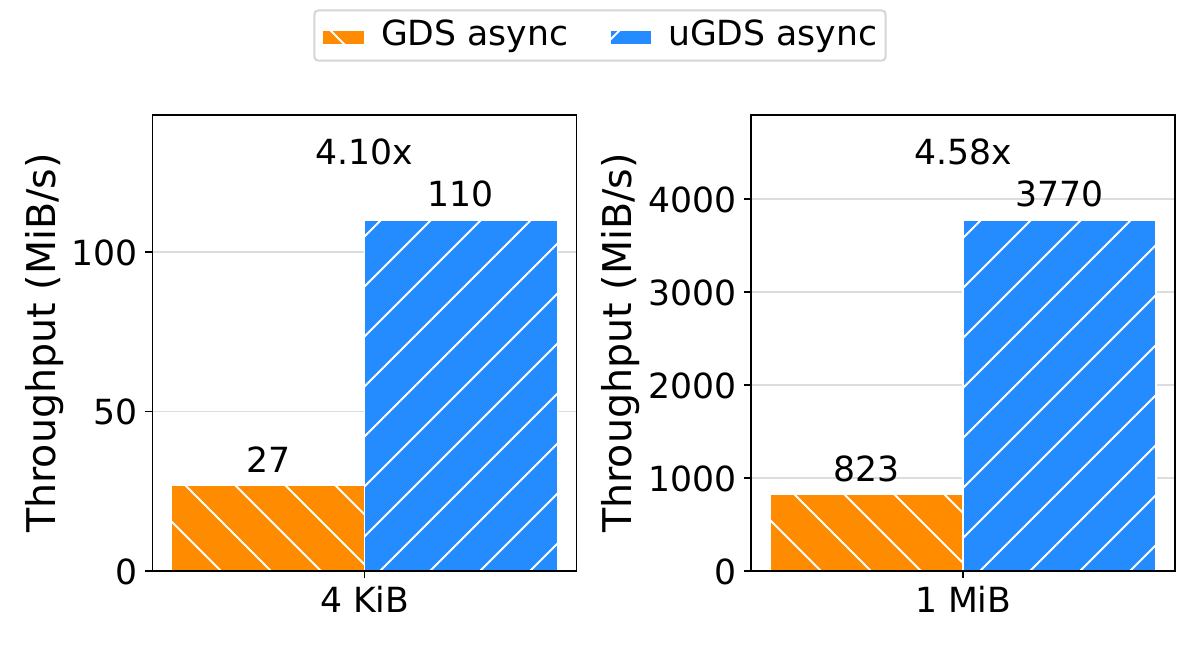}
  \captionof{figure}{Stream-ordered read throughput with synchronization after each
  request. Each bar is the median of three process runs.}
  \label{fig:eval-async}

  \vspace{2pt}

  \includegraphics[width=\columnwidth]{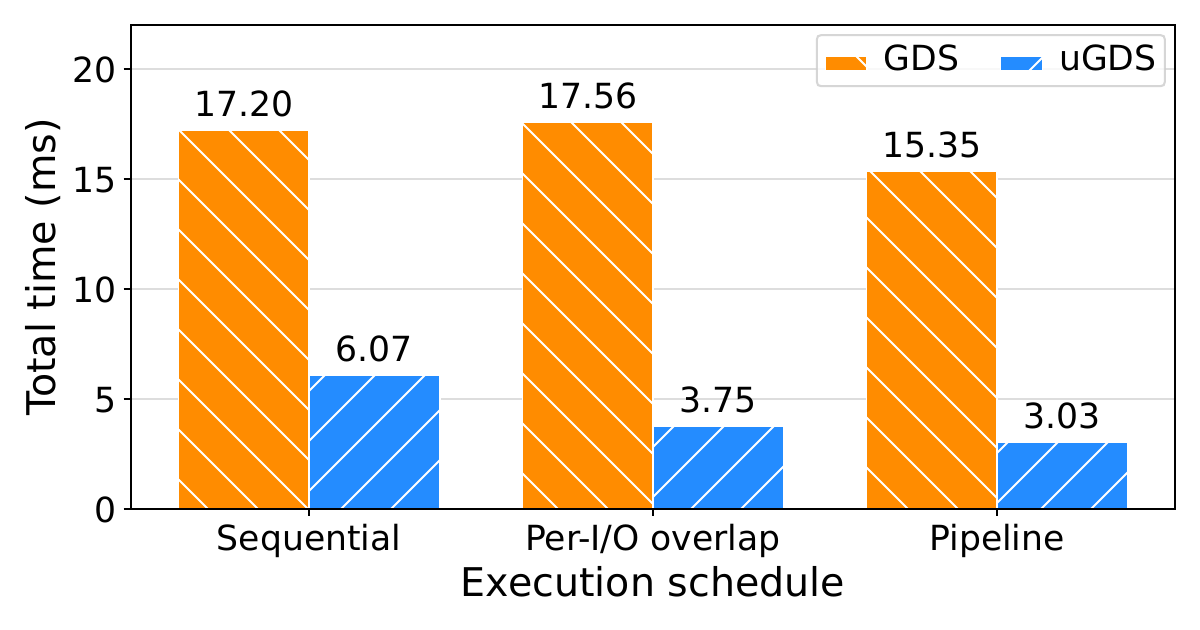}
  \captionof{figure}{Median total time for 16 1-MiB writes and identical GPU compute
  under three synchronization schedules; lower is better.}
  \label{fig:eval-overlap}
\end{figure}

\begin{figure}[!t]
  \centering
  \includegraphics[width=\columnwidth]{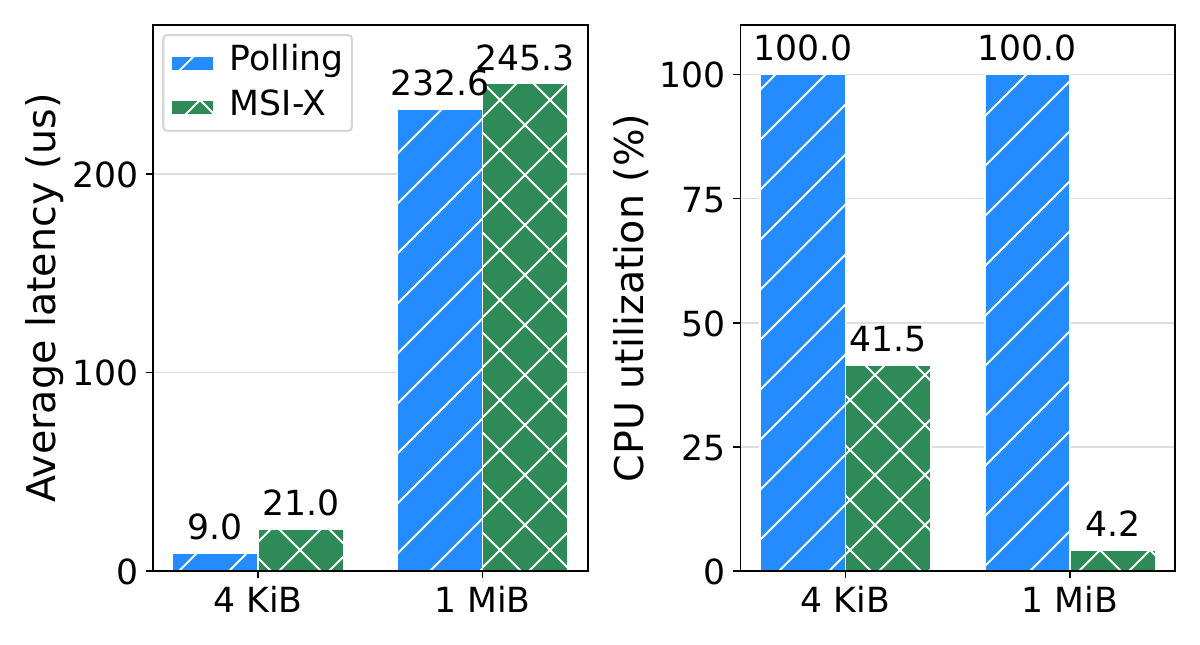}
  \caption{uGDS completion-policy tradeoff for one issuing thread and one QP.
  CPU utilization is issuing-thread CPU time divided by wall time.}
  \label{fig:eval-completion}
\end{figure}

\subsection{Stream-Ordered Asynchronous I/O and Pipeline Overlap}
\label{sec:eval-async}

Stream ordering allows applications to express dependencies between storage
transfers and GPU kernels. We evaluate two aspects: the cost of an individual
stream-ordered read followed by synchronization, and the benefit of scheduling
I/O and computation with fewer intermediate synchronization points. Both uGDS
and GDS expose stream-ordered APIs; the comparison therefore measures how
efficiently each backend advances the same stream schedule.

\paragraph{Serialized stream execution}
To isolate stream dispatch overhead, Figure~\ref{fig:eval-async} evaluates an
intentionally serialized loop where each asynchronous read is immediately
followed by \texttt{cudaStreamSynchronize}, forcing dispatch, hardware DMA, and
completion waiting directly onto the critical path.

uGDS achieves 110\,MiB/s at 4\,KiB and 3770\,MiB/s at 1\,MiB, outperforming GDS
(27 and 823\,MiB/s) by 4.10$\times$ and 4.58$\times$, respectively. This gap
combines the latency of stream dispatch, the storage operation, and completion
waiting. At 4\,KiB, the median of the three per-run mean request latencies is
35.43\,$\mu$s for uGDS and 145.46\,$\mu$s for GDS.

The synchronous measurements in Section~\ref{sec:eval-sync} identify the
underlying storage-path contribution. With two issuing threads and 4-KiB reads,
uGDS reduces the median per-run mean latency from 62.21 to 11.56\,$\mu$s
(5.38$\times$). The asynchronous uGDS entry point preserves this data path:
it enqueues a \texttt{cudaLaunchHostFunc} callback, and the callback invokes the
same \texttt{do\_io\_internal()} engine used by \texttt{uGDSRead()}. That engine
formats the NVMe command, rings a mapped doorbell, and polls the mapped completion
queue in user space. Consequently, the asynchronous interface adds stream
ordering around the low-latency synchronous engine while allowing the issuing
thread to return after enqueue. The serialized result includes the additional
callback scheduling and stream-synchronization costs, but retains the latency
advantage of direct user-space command submission.

\paragraph{Overlap and pipelining}
Figure~\ref{fig:eval-overlap} holds the workload fixed at sixteen 1-MiB
writes and identical GPU computation, then varies the synchronization schedule.
Sequential execution completes each I/O and compute stage before proceeding.
Per-I/O overlap uses separate streams for the paired stages but waits after
each pair. Pipelined execution enqueues all pairs before the final barrier.

Under these execution schedules, uGDS reduces total time from 6.07\,ms
(Sequential) to 3.75\,ms (Per-I/O overlap) and 3.03\,ms (Pipelining), for an
aggregate $2.00\times$ speedup. Stream dependencies permit storage and compute
work from different pairs to overlap, while deferring the final barrier until
all 16 pairs are queued reduces latency by a further 19.2\% relative to the
per-I/O schedule. The callback blocks until its synchronous user-space request
completes, so the low service time of \texttt{do\_io\_internal()} lets the I/O
stream advance promptly and exposes a useful overlap window to the compute
stream.

GDS records 17.20, 17.56, and 15.35\,ms under the same three schedules. Its
stream API expresses the same dependencies, yet the measured asynchronous path
advances requests more slowly: per-I/O overlap provides no reduction, and full
pipelining improves on the sequential baseline by 10.8\%. Because each storage
submission precedes the corresponding kernel launch in the benchmark loop,
submission backpressure also delays the host from filling the compute stream.
The combination of uGDS's quickly enqueued host callback and its low-latency
user-space synchronous engine keeps both streams supplied, leaving uGDS
$5.07\times$ faster under full pipelining.
Section~\ref{sec:eval-lmcache} further examines this submission backpressure
within real-world application pipelines.

\subsection{Completion Policy: Polling vs.\ MSI-X Interrupts}
\label{sec:eval-policy}

To guide system deployment, Figure~\ref{fig:eval-completion} 
quantitatively analyzes the tradeoff between completion latency and 
host CPU consumption across busy-polling and MSI-X interrupt-driven 
policies.

\paragraph{Small I/O ($<$16\,KiB): Polling Dominates} For 4-KiB 
requests, user-space polling slashes latency by 57.2\% (from 21.0\,$\mu$s 
down to 8.97\,$\mu$s), boosting throughput by 2.33$\times$ at the expense 
of pinning a CPU core at 100\% utilization. In this regime, 
hardware transfer time is so short that kernel interrupt routing and 
context-switching overheads in \texttt{ppoll()} account for over half of 
total request latency.
\paragraph{Large I/O ($\ge$1\,MiB): MSI-X Relieves CPU at Negligible Cost} 
Conversely, for 1-MiB transfers, the SSD transfer time dominates the end-to-end 
latency. Here, MSI-X mode reduces issuing-thread CPU utilization from 
100\% down to just 4.15\% (a 95.8\% compute saving) while incurring a 
negligible 5.5\% latency penalty (245.3\,$\mu$s vs.\ 232.6\,$\mu$s).

Busy polling is ideal for low-latency, dedicated inference nodes where microsecond 
latency directly affects prompt responsiveness. Meanwhile, our kernel-assisted 
MSI-X mode provides an exceptionally power- and core-efficient solution for throughput-bound 
analytics pipelines where host CPU cycles are shared with complex CPU preprocessing.

\begin{figure*}[!t]
  \centering
  \begin{minipage}[t]{0.48\textwidth}
    \centering
    \includegraphics[width=\linewidth]{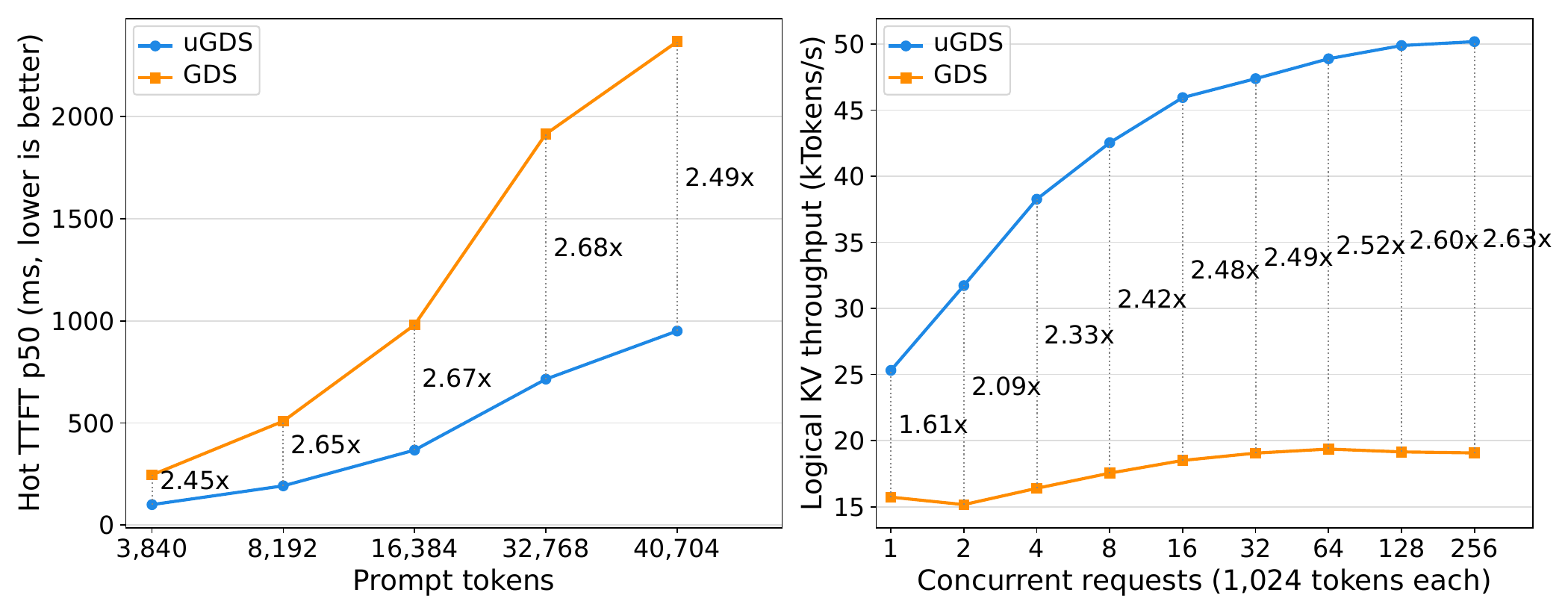}
    \captionof{figure}{LMCache SSD-only cache-hit performance with Qwen3-0.6B
    on an A100. Left: median TTFT as prompt length increases (lower is better).
    Right: logical KV throughput for 1,024-token requests as concurrency
    increases (higher is better). Annotations report the corresponding speedups.}
    \label{fig:lmcache-e2e}
  \end{minipage}\hfill
  \begin{minipage}[t]{0.48\textwidth}
    \centering
    \includegraphics[width=\linewidth]{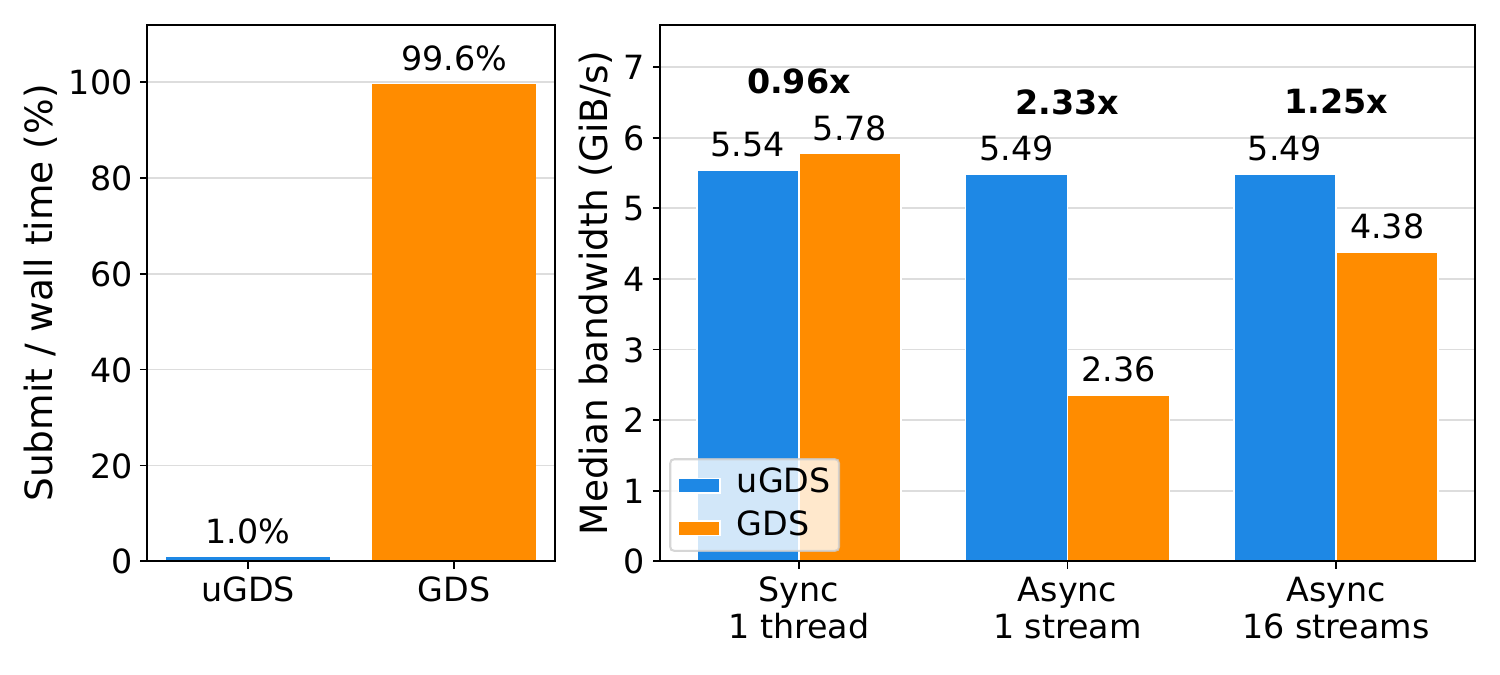}
    \captionof{figure}{LMCache storage-path breakdown. Left: caller-side
    submission time as a fraction of elapsed time at asynchronous depth 64
    through GDSContext. Right: median bandwidth for direct 16-MiB backend reads
    under synchronous and asynchronous execution.}
    \label{fig:lmcache-breakdown}
  \end{minipage}
\end{figure*}

\subsection{LMCache Integration and End-to-End Evaluation}
\label{sec:eval-lmcache}

In long-context LLM serving, reusing precomputed KV caches is essential to
avoid redundant prompt precomputation and slash time-to-first-token (TTFT)
latency. Emerging caching architectures (e.g., LMCache~\cite{lmcache} and
Mooncake~\cite{mooncake}) increasingly adopt NVMe SSDs as an elastic,
cost-effective local tier to retain massive KV states beyond GPU memory limits.
In such tiered serving systems, the underlying storage backend's retrieval
throughput and dispatch overhead directly govern end-to-end serving goodput and
latency during cache hits.

To evaluate real-world application impact, we benchmark uGDS as the local NVMe
storage backend for LMCache~\cite{lmcache}. We implement a drop-in uGDS adapter
that adheres to LMCache's stream-ordered asynchronous backend interface and has
been integrated into the codebase~\cite{lmcache-ugds}.
Table~\ref{tab:lmcache-configuration} summarizes the deployment environment.
To strictly isolate the storage subsystem's contribution, we disable vLLM~\cite{vllm}'s
in-memory automatic prefix caching and enforce SSD-only cache hits, ensuring
that all reusable KV blocks are retrieved directly across the evaluated storage
paths.

\begin{table}[!t]
  \caption{LMCache end-to-end experimental configuration.}
  \label{tab:lmcache-configuration}
  \centering
  \small
  \begin{tabular}{@{}p{0.27\columnwidth}p{0.64\columnwidth}@{}}
    \toprule
    Component & Configuration \\
    \midrule
    GPU & NVIDIA A100-SXM4-40GB \\
    Storage & Samsung 990 PRO, PCIe Gen4 $\times$4; cross-root-port P2P \\
    Model & Qwen3-0.6B~\cite{qwen3} \\
    vLLM~\cite{vllm} & 0.20.1+cu129 \\
    LMCache & 0.5.1.dev35 \\
    PyTorch & 2.11.0+cu129 \\
    CUDA & 12.9 \\
    \bottomrule
  \end{tabular}
\end{table}

\paragraph{End-to-End Impact: TTFT Scaling and Serving Goodput}
We first evaluate single-request responsiveness across scaling context lengths
(Figure~\ref{fig:lmcache-e2e}, left) alongside multi-tenant serving capacity
under concurrent load (Figure~\ref{fig:lmcache-e2e}, right).

As the prompt length expands from 3,840 to 40,704 tokens, uGDS consistently
reduces median TTFT by 2.45$\times$--2.68$\times$. At 32,768 tokens, uGDS slashes
TTFT from 1915\,ms to 715\,ms, achieving a 62.7\% latency reduction. Because
longer cached prefixes require transferring proportionally larger KV states
before attention computation can begin, this stable speedup confirms that
reducing low-level storage latency directly and linearly accelerates serving
responsiveness.

Under multi-tenant workloads with 1,024-token prompts, this latency advantage
translates directly into higher serving goodput. As request concurrency scales
from 1 to 256, GDS saturates prematurely near 19\,kTokens/s by concurrency 32,
hobbled by cumulative request queueing in its storage path. In contrast, uGDS
scales smoothly to 50.2\,kTokens/s at concurrency 256, delivering a
2.63$\times$ goodput improvement. This demonstrates that uGDS substantially
shifts the saturation boundary of the inference server outward, enabling high
cache-hit traffic bursts without inducing serving pipeline collapse.

\paragraph{Root Cause: Profiling Latency Breakdown and Submission Backpressure}
Because both backends operate over identical NVMe hardware, we dissect the KV
restoration pipeline at concurrency 256 to pinpoint the underlying performance
bottleneck. Restoring a cached KV block comprises two stages: (1) retrieving raw
tensors from NVMe into a staging buffer via P2P DMA, and (2) scattering chunked
tensors into vLLM's paged GPU memory pool. Profiling reveals that GPU-side KV
scattering consumes identical time between both backends (0.189\,ms for uGDS vs.\
0.193\,ms for GDS), confirming that GPU compute and memory layout transformations
are not the limiter. Instead, median SSD-to-buffer transfer latency drops from
56.04\,ms under GDS to 19.98\,ms under uGDS (a 2.80$\times$ improvement), proving
that application-level gains originate entirely from the storage subsystem.

To isolate why GDS's I/O path degrades under load,
Figure~\ref{fig:lmcache-breakdown} (left) profiles caller-side submission overhead
inside LMCache's \texttt{GDSContext} wrapper at asynchronous queue depth 64. Each
1,024-token transaction moves 112\,MiB in four 28-MiB chunks, which are split
into 16-MiB and 12-MiB transfers at buffer registration boundaries. Under GDS,
time spent blocked inside submission calls accounts for an overwhelming 99.6\%
of wall-clock time, exposing severe submission backpressure and kernel-transition
stalls. Conversely, uGDS spends only 1.0\% of elapsed time in submission, rapidly
delegating transfers to user-space host callbacks that format SQEs and poll CQs
asynchronously on background worker threads.

Figure~\ref{fig:lmcache-breakdown} (right) validates this diagnosis by
measuring direct 16-MiB reads outside the wrapper. When executed synchronously
with a single thread, both backends achieve comparable wire rates: 5.54\,GiB/s
for uGDS and 5.78\,GiB/s for GDS. However, under a single asynchronous stream, the
exact operational model used in LMCache, GDS's effective throughput collapses to
2.36\,GiB/s (retaining only 40.8\% of its synchronous speed). In contrast, uGDS
sustains 5.49\,GiB/s (98.9\% retention), outperforming GDS by 2.33$\times$.
While expanding GDS to 16 concurrent streams partially mitigates this issue
(recovering to 4.38\,GiB/s), uGDS delivers saturation bandwidth out of the box
on a single stream without requiring complex multi-stream orchestration.
In summary, GDS's end-to-end performance degradation in LMCache stems directly
from heavy synchronous submission blocking and inefficient single-stream
asynchronous dispatch, both of which are fundamentally resolved by uGDS's
user-space non-blocking callback architecture.

\FloatBarrier
\section{Conclusion and Outlook}
\begingroup
\setlength{\emergencystretch}{1em}

This paper presented uGDS, an open-source user-space GPU Direct Storage
library. uGDS keeps the entire NVMe command lifecycle in user space and limits
the kernel to one-time privileged setup; it normalizes CUDA and HIP memory into
one controller-visible abstraction; and because it owns the hardware queues, it
partitions them by traffic class and selects the completion policy per I/O
mode. The result is a clean division of responsibilities: the CPU
orchestrates, the SSD transfers, and the GPU computes. Our results show that
CPU-initiated GPU direct I/O, carefully engineered on a multicore host,
captures the user-space benefit without consuming GPU compute or changing how
applications manage data. uGDS reduces small-request latency by up to 84.1\%,
outperforms GDS in every evaluated batch configuration, and sustains device
bandwidth on a single stream; in LMCache, it reduces median TTFT by up to
62.7\% and raises serving goodput by 2.63$\times$. Because the full stack is
open source, users can trace these behaviors, debug anomalies, and tune
policies directly instead of inferring them from a closed library.

uGDS is under active development. Our roadmap focuses on three directions.

\paragraph{NVMe-oF and Multi-SSD Support}
Each uGDS handle currently drives a single local PCIe SSD. Within a host, we
plan to aggregate multiple SSDs behind one handle and stripe requests across
them automatically, steering each GPU toward the SSDs that share its PCIe
switch or root port so that peer-to-peer traffic avoids the CPU root complex
where possible. Across hosts, we plan to support NVMe over Fabrics (NVMe-oF)
targets, moving data over RDMA directly between remote SSDs and local GPU
memory. The DMA-buf export of registered GPU buffers
(Section~\ref{sec:gpu-memory-abstraction}) already provides the memory
registration an RDMA NIC requires. Both extensions reuse the same user-space
queue engine and API, so applications need not change when data moves from a
local SSD to a striped array or a remote target.

\paragraph{A Lightweight Built-In Scheduler}
Synchronous requests are currently dispatched round-robin across QPs, and each
request holds its QP until its final completion
(Section~\ref{sec:queue-scheduling}). This keeps the fast path simple, but it
ignores queue occupancy, and a large request can delay small requests that
land on the same QP. We plan to add a simple scheduler between the API and the
hardware queues. It will track outstanding work per QP, place each request on
the least-loaded queue, split large requests into chunks that can interleave
with small ones, and bound per-queue depth. It will also support
ordering-aware stream batches: a group of requests is ordered as a whole with
respect to its stream, while requests inside the group carry no mutual order
and can be issued concurrently on a deep queue and completed out of order.
This matches workloads such as KV-cache restoration, which need ordering only
between a batch of reads and the kernel that consumes them, without forcing
applications to fan I/O out across streams. The scheduler will stay in user
space and avoid shared locks on the common path, so it adds little to the
per-request cost that the current design minimizes. It will also expose
per-queue statistics, making scheduling decisions observable in the same way
as the rest of the command lifecycle.

\paragraph{Built-In Scheduling Modes}
As Section~\ref{sec:eval-policy} shows, workloads want different tradeoffs from
the same hardware. On top of the scheduler, we plan to provide a small set of
built-in modes that applications select per handle or per stream, instead of
hand-tuning queue counts and completion policies. A \emph{latency} mode will
use busy polling and shallow queues for small critical-path reads such as
KV-cache loads. A \emph{throughput} mode will use deep queues, doorbell
coalescing, and interrupt-driven completion to sustain bandwidth while
freeing CPU cores. A \emph{priority} mode will serve latency-critical requests
ahead of background traffic such as KV-cache offload writes and checkpoints.
An \emph{adaptive} mode will switch between polling and interrupts based on
observed request size and load. Because the library is open source, users can
also implement their own modes against the same scheduler interface.
\par\endgroup

\section{Acknowledgments}

We thank all GitHub contributors for their valuable contributions and feedback, which helped improve the quality of this work. We also thank the members of the ScaleX Lab for their valuable comments and discussions. This research was supported by HKUST (GZ) startup funding.

\end{document}